\documentclass[preprint,12pt,authoryear]{elsarticle}
\usepackage{amsmath,mathrsfs,amssymb,wasysym}
\journal{EPSL}

\usepackage{fullpage}
\usepackage{lineno}
\usepackage{xcolor}
\usepackage{mathtools}
\usepackage{ragged2e}
\usepackage{array,longtable,tabularx}
\usepackage{adjustbox}
\usepackage{appendix}
\usepackage{upgreek}
\usepackage{url}

\usepackage[flushleft]{threeparttable}

\begin{document}

\begin{frontmatter}
\title{Quantification of classical and non-classical crystallization pathways in calcite precipitation}
\author[1]{Zhongtian Zhang}
\author[2]{Jiuyuan Wang}
\affiliation[1]{organization={Earth and Planets Laboratory, Carnegie Institution for Science},
city={Washington},
state={DC},
country={USA}}
\affiliation[2]{organization={School of Earth and Space Sciences, Peking University},
city={Haidian},
state={Beijing},
country={China}}

\begin{abstract}
Crystal precipitation from aqueous solution occurs through multiple pathways. Besides the classical ion-by-ion addition, non-classical crystallization mechanisms, such as multi-ion polymer and nano-particle attachment, could be significant. These non-classical crystallization processes have been observed with advanced microscopy, yet detailed quantification of their contributions remains challenging. Building from paired Ca and Sr isotope observations, we develop a theoretical framework to quantify the contributions of classical and non-classical crystallization pathways on the precipitation of the calcium carbonate mineral calcite, a common precipitate in nature. We demonstrate that the classical crystallization pathway alone is insufficient to account for the observed isotope behaviors and, thus, the entire calcite precipitation process. We further present a surface kinetic model that incorporates non-classical crystallization pathways. This model enables the characterization of the roles of classical and non-classical crystallization mechanisms in calcite precipitation. The results suggest that the relative contribution of non-classical crystallization pathways increases with saturation state and can, under high supersaturation levels, be comparable to or greater than the classical pathway. The presented theoretical framework readily explains observed trace element partitioning and isotope fractionation behaviors during calcite precipitation and can be further expanded onto other mineral systems to gain insights into crystal growth mechanisms.
\end{abstract}


\end{frontmatter}

\section{Introduction}  \label{sec:intro}
The precipitation of crystals from aqueous solutions is an important subject in geological and environmental sciences. The classical theory treats crystallization as a process of ion-by-ion attachment, where ions attach at available kink sites along ledges onto crystal surface \citep[e.g.,][]{burton1951growth}. Recently, numerous studies have presented evidence for non-classical crystallization pathways, where larger species ranging from polymeric multi-ion complexes to nano-particles attach directly onto crystal surfaces, and found that they occur simultaneously with the classical crystallization pathway \citep[Fig.~\ref{fig:cartoon}; e.g.,][]{nielsen2014situ, de2015crystallization}. Despite the existence of direct microscopic observations on classical and non-classical crystallization mechanisms \citep[e.g.,][]{li2012direction, lupulescu2014situ, putnis2021crystallization}, quantifying their contributions under different conditions remains challenging \citep[e.g.,][]{de2015crystallization, ivanov2014oriented, putnis2021crystallization}. 

The partitioning of trace elements and the fractionation of stable isotopes during calcite precipitation are strongly affected by crystallization kinetics \citep[e.g.,][]{watkins2017kinetic}. Trace element partitioning during calcite growth has been studied extensively in both natural and laboratory settings \citep[e.g.,][]{lorens1981sr, carpenter1992srmg, paquette1995relationship, nehrke2007dependence, tang2008I, gabitov2006partitioning, gabitov2014crystal}, and several theoretical models have been developed to explain these observations \citep{depaolo2011surface, nielsen2012self, nielsen2013general, jia2022model}. Advances in stable isotope analyses of carbonate-incorporated major and trace metals (e.g., Ca, Li, Mg, Sr, Ba) \citep[e.g.,][]{tang2008II, bohm2012strontium, mavromatis2013kinetics, mavromatis2020experimental, zhang2020equilibrium, fuger2022effect, alkhatib2017calcium} could provide new insights to calcite growth kinetics and crystallization pathways. While most existing models are based on the classical crystallization pathway \citep{depaolo2011surface, nielsen2012self, nielsen2013general, jia2022model}, these new observations highlight a more complex and diverse range of carbonate precipitation processes and thus offer additional constraints on previous precipitation models. In this study, with the paired observations of Ca and Sr isotope fractionations \citep[e.g.,][]{tang2008II, tang2008I, bohm2012strontium, wang2021stable}, we demonstrate the inadequacies of previous models to account for the full range of the observed calcite precipitation processes, and then provide new additions to incorporate the non-classical crystallization mechanisms. Applying this new framework, we quantify the roles of classical and non-classical crystallization mechanisms at different precipitation rates and supersaturation levels. This model framework can also be applied to other crystal systems and tested with other paired element and isotope measurements.

\section{A reassessment of previous models on calcite precipitation}  \label{sec:previous}
For calcite precipitated from aqueous solutions, the isotopic compositions of the main block-building element, Ca, and its trace element substitution, Sr, depend on the rate of precipitation \citep[e.g.,][]{fietzke2006determination, fantle2007isotopes, jacobson2008delta44ca, tang2008II, tang2008I, bohm2012strontium, alkhatib2017calcium, shao2021impact}. The Ca and Sr isotope fractionations are described in the $\Delta$-notation as
\begin{linenomath} \begin{equation}
    \mathrm{\Delta^{44/40}Ca} = 1000\permil \times \left[\frac{(\mathrm{^{44}Ca}/\mathrm{^{40}Ca})_\mathrm{cal}}{(\mathrm{^{44}Ca}/\mathrm{^{40}Ca})_\mathrm{aq}}-1\right],    
\end{equation}\end{linenomath} 
\begin{linenomath} \begin{equation}
    \mathrm{\Delta^{88/86}Sr} = 1000\permil \times \left[\frac{(\mathrm{^{88}Sr}/\mathrm{^{86}Sr})_\mathrm{cal}}{(\mathrm{^{88}Sr}/\mathrm{^{86}Sr})_\mathrm{aq}}-1\right],    
\end{equation}\end{linenomath} 
where $(\mathrm{^{44}Ca}/\mathrm{^{40}Ca})_\mathrm{cal}$, $(\mathrm{^{88}Sr}/\mathrm{^{86}Sr})_\mathrm{cal}$, $(\mathrm{^{44}Ca}/\mathrm{^{40}Ca})_\mathrm{aq}$, and  $(\mathrm{^{88}Sr}/\mathrm{^{86}Sr})_\mathrm{aq}$ are the Ca and Sr isotope ratios of the calcite crystal and the aqueous solution, respectively. The Sr/Ca elemental partitioning is described by the partition coefficient,
\begin{linenomath} \begin{equation}
    K = \frac{(\mathrm{Sr}/\mathrm{Ca})_\mathrm{cal}}{(\mathrm{Sr}/\mathrm{Ca})_\mathrm{aq}},
\end{equation}\end{linenomath} 
where $(\mathrm{Sr}/\mathrm{Ca})_\mathrm{cal}$ and $(\mathrm{Sr}/\mathrm{Ca})_\mathrm{aq}$ are the $\mathrm{Sr}/\mathrm{Ca}$ ratios of the calcite crystal and the aqueous solution. At low precipitation rates, $\mathrm{\Delta^{44/40}Ca}$, $\mathrm{\Delta^{88/86}Sr}$, and $K$ converge to their equilibrium values, $\mathrm{\Delta^{44/40}Ca_{eq}}$, $\mathrm{\Delta^{88/86}Sr_{eq}}$, and $K_\mathrm{eq}$ \citep[e.g.,][]{fantle2007isotopes, jacobson2008delta44ca, bohm2012strontium,zhang2020equilibrium}; at sufficiently high precipitation rates, they are usually assumed to converge to their kinetic limits, $\mathrm{\Delta^{44/40}Ca_{inf}}$, $\mathrm{\Delta^{88/86}Sr_{inf}}$, and $K_\mathrm{inf}$ \citep[e.g.,][]{depaolo2011surface, nielsen2012self, nielsen2013general, jia2022model}. With natural and experimental observations of calcite precipitated at different rates, we suggest equilibrium- and kinetic-limit parameters listed in Tab.~\ref{Tab:parameters} (also see \ref{app:limit_parameters} for detailed discussion).

\cite{watson2004conceptual} invoked the ``growth entrapment'' model to explain the variation of trace element partitioning and isotope fractionation with precipitation rate. This model assumes that the surface of the growing crystal is in equilibrium with the aqueous solution. However, in order to achieve surface equilibrium, the net precipitation rate must be substantially less than the rate of ion detachment, which is hard to reconcile with observations from controlled growth experiments \citep[as pointed out by][]{depaolo2011surface}. This implies that the crystal surface and aqueous solution are rarely in equilibrium, and the kinetics of surface reaction plays an important role. For this reason, we focus on surface kinetic models in the following text.

\subsection{D11 model}
\cite{depaolo2011surface} developed a surface reaction model for kinetic processes of trace element partitioning and isotope fractionation during calcite precipitation from aqueous solutions, which is  the first-of-its-kind to account for most of the experimental observations. In this model (hereafter referred to as the ``D11'' model), precipitation is considered as the overall result of the forward reaction (i.e., ion attachment from the aqueous solution onto the crystal surface) and the backward reaction (i.e., ion detachment from the crystal surface). The forward and backward reactions are assumed to be associated with constant levels of isotope fractionation and trace element partitioning, and the net fractionation or partitioning behaviors vary with precipitation rate due to the competition between backward and forward reactions. In this framework, $\mathrm{\Delta^{44/40}Ca}$, $\mathrm{\Delta^{88/86}Sr}$, and $K$ can all be expressed as functions of the forward-to-backward reaction rate ratio, $R_\mathrm{f}/R_\mathrm{b}$ (here $R_\mathrm{f}$ and $R_\mathrm{b}$ are the forward and backward reaction rates). The derivations of these functions are provided in \ref{app:D11}.

In precipitation experiments, $\mathrm{\Delta^{44/40}Ca}$, $\mathrm{\Delta^{88/86}Sr}$, and $K$ can be directly determined. The net calcite precipitation rate, $R_\mathrm{p} = R_\mathrm{f}-R_\mathrm{b}$, can be also obtained, but direct measurements of $R_\mathrm{f}$ and $R_\mathrm{b}$ remain elusive. A straightforward way to test a kinetic model is to evaluate how well it explains the variations of trace element partitioning and isotope fractionation behaviors with precipitation rate. However, this would involve assumptions for the conversion between the model variable $R_\mathrm{f}/R_\mathrm{b}$ and the experimental observable $R_\mathrm{p}$. In \cite{depaolo2011surface}, two models were described: ``Model 1'', where the backward reaction rate $R_\mathrm{b}$ is held constant, and ``Model 2'', where $R_\mathrm{b}$ is a function of $R_\mathrm{p}$. The specific choice of the $R_\mathrm{b}$--$R_\mathrm{p}$ relation significantly impacts the predicted relations and, as a result, ``Model 2'' yields better viability than `Model 1''. As detailed in \ref{app:D11M2}, the flexibility in adjusting the $R_\mathrm{b}$--$R_\mathrm{p}$ relation often allows for a good fit to individual observation, such as $\mathrm{\Delta^{44/40}Ca}$--$R_\mathrm{p}$, $K$--$R_\mathrm{p}$, and even $\mathrm{\Delta^{88/86}Sr}$--$R_\mathrm{p}$ (which was unattainable at the time of D11 model development). With recent advancements in Sr isotope measurements, a more comprehensive assessment of a model lies in its capability to concurrently account for the variations of $\mathrm{\Delta^{44/40}Ca}$, $\mathrm{\Delta^{88/86}Sr}$, and $K$ with $R_\mathrm{p}$, and equivalently their interrelations \citep{bohm2012strontium}. Since $\mathrm{\Delta^{44/40}Ca}$, $\mathrm{\Delta^{88/86}Sr}$, and $K$ are all functions of $R_\mathrm{f}/R_\mathrm{b}$ under the D11 framework, we can reformulate them as functions of each other and eliminate $R_\mathrm{f}/R_\mathrm{b}$ to predict their mutual correlations. This approach circumvents any assumptions regrading the $R_\mathrm{b}$--$R_\mathrm{p}$ relation, such that the choice between `Model 1'' and `Model 2'' is inconsequential. In the following text, we outline this new evaluation of the D11 model regarding the correlations of $K$, $\mathrm{\Delta^{88/86}Sr}$, and $\mathrm{\Delta^{44/40}Ca}$.

Under the D11 model framework, with given Ca isotope fractionation factor ($\mathrm{\Delta^{44/40}Ca}$), the Sr/Ca partition coefficient ($K$) and Sr isotope fractionation factor ($\mathrm{\Delta^{88/86}Sr}$) can be predicted via the following relations (see \ref{app:D11} for detailed derivations),
\begin{linenomath} 
\begin{equation}
    K_\mathrm{D11} = K_\mathrm{inf} + \left(K_\mathrm{eq}-K_\mathrm{inf}\right)
    \times \left(\frac{K_\mathrm{inf}}{K_\mathrm{eq}}\right) \left(  \frac{ \mathrm{\Delta^{44/40}Ca_{eq}}-\mathrm{\Delta^{44/40}Ca_{M}} }{ \mathrm{\Delta^{44/40}Ca_{M}}-\mathrm{\Delta^{44/40}Ca_{inf}} } + \frac{K_\mathrm{inf}}{K_\mathrm{eq}} \right)^{-1},  \label{KSr-DELTA_Ca-D11}  
\end{equation}
\begin{multline}
    \mathrm{\Delta^{88/86}Sr_\mathrm{D11}} = \mathrm{\Delta^{88/86}Sr_{inf}} + \left(\mathrm{\Delta^{88/86}Sr_{eq}}-\mathrm{\Delta^{88/86}Sr_{inf}}\right)  \\
    \times \left( \frac{K_\mathrm{inf}}{K_\mathrm{eq}} \right) \left( \frac{ \mathrm{\Delta^{44/40}Ca_{eq}}-\mathrm{\Delta^{44/40}Ca_{M}} }{ \mathrm{\Delta^{44/40}Ca_{M}}-\mathrm{\Delta^{44/40}Ca_{inf}} } + \frac{K_\mathrm{inf}}{K_\mathrm{eq}} \right)^{-1},   \label{DELTA_Sr-DELTA_Ca-D11}  
\end{multline}
\end{linenomath}
where $K_\mathrm{D11}$ and $\mathrm{\Delta^{88/86}Sr_\mathrm{D11}}$ are the $K$ and $\mathrm{\Delta^{88/86}Sr}$ values predicted under D11 framework, and $\mathrm{\Delta^{44/40}Ca_{M}}$ is the experimentally measured $\mathrm{\Delta^{44/40}Ca}$. With our preferred parameters for the equilibrium and kinetic limits, the D11 predictions cannot satisfactorily match the observed $K$--$\mathrm{\Delta^{44/40}Ca}$ and $\mathrm{\Delta^{88/86}Sr}$--$\mathrm{\Delta^{44/40}Ca}$ correlations (Figs.~\ref{fig:previous_models}a,b). Specifically, for $\mathrm{\Delta^{88/86}Sr}$, the root-square mean error between the prediction of Eq.~\ref{DELTA_Sr-DELTA_Ca-D11} and the experimental data is ${\sim}0.12\permil$, which is much larger than the analytical uncertainty \citep[${\sim}0.02\permil$;][]{bohm2012strontium, wang2023investigation}.

We note that adjusting the equilibrium parameters (i.e., $\mathrm{\Delta^{44/40}Ca_{eq}}$, $K_\mathrm{eq}$, and $\mathrm{\Delta^{88/86}Sr_{eq}}$) may result in improved fittings of the controlled experiment data. For example, if we assume $K_\mathrm{eq} = 0.07$ \citep{depaolo2011surface}, the D11 prediction on $K$--$\mathrm{\Delta^{44/40}Ca}$ relation would match better with the experimental calcite data. However, these fittings would not align with the ocean crust calcite precipitated at lower rates (e.g., the open circles in Fig.~\ref{fig:previous_models}, which were not available when \cite{depaolo2011surface} was published). In fact, the equilibrium parameters have been independently constrained by measurements of samples in natural systems. The equilibrium Ca isotope fractionation factor has been inferred to be zero ($\mathrm{\Delta^{44/40}Ca_{eq}}=0\permil$) in both slowly-precipitated deep sea sediments \citep[precipitation rate of $10^{-18}$--$10^{-17}\ \mathrm{mol/m^2/s}$;][]{fantle2007isotopes} and carbonate aquifer \citep[][]{jacobson2008delta44ca}, which has recently been confirmed in a laboratory controlled experiment \citep[][]{harrison2023equilibrium}. The equilibrium Sr/Ca partition coefficient was inferred from deep-sea sediments and pore fluids to be $K_\mathrm{eq}=0.025$ at 25$^{\circ}$C \citep[][]{zhang2020equilibrium}. Thus, we suggest that $\mathrm{\Delta^{44/40}Ca_{eq}}$ and $K_\mathrm{eq}$ are robustly constrained by these observations and should be fixed in the model. Modifying these equilibrium parameters to improve the fit to experimental calcite measurements would compromise the model's ability to replicate natural observations of slowly precipitated calcite. 

For $\mathrm{\Delta^{88/86}Sr}$, \cite{bohm2012strontium} reported values as small as $-0.05\permil$ for ocean crust calcite. The inferred $\mathrm{\Delta^{44/40}Ca}$ and $K$ of these samples \citep{bohm2012strontium} are still higher in magnitude than the deep sea sediment \citep[][]{fantle2007isotopes, zhang2020equilibrium}; thus, the actual $\mathrm{\Delta^{88/86}Sr_{eq}}$ must be less negative than $-0.05\permil$ and is likely close to $0\permil$ \citep[][also see our \ref{app:limit_parameters}]{bohm2012strontium}. Therefore, we suggest $\mathrm{\Delta^{88/86}Sr_{eq}} = 0\permil$ and apply it as the preferred value throughout this paper. Despite this preference, we acknowledge that direct observations only confirm $\mathrm{\Delta^{88/86}Sr_{eq}}$ to be smaller than $-0.05\permil$ in magnitude. Given this constraint, we also conduct calculations with $\mathrm{\Delta^{88/86}Sr_{eq}} = -0.05\permil$ and show through the results (Fig.~\ref{fig:previous_models}c) that this adjustment does not impact our conclusion. In \ref{app:parameters}, we also demonstrate that the discrepancy between observations and D11 model predictions cannot be resolved by varying parameters such as $K_\mathrm{inf}$, $\mathrm{\Delta^{44/40}Ca_{inf}}$, and $\mathrm{\Delta^{88/86}Sr_{inf}}$ within a reasonable range. Thus, the conundrum introduced by new $\mathrm{\Delta^{88/86}Sr}$ observations cannot be resolved by adjusting parameters; additional mechanisms are required.

\subsection{Ion-by-Ion model}
An updated version of the surface kinetic model for trace element partitioning and isotope fractionation was developed by \cite{nielsen2012self, nielsen2013general}. This model (hereafter referred to as the ``Ion-by-Ion'' model) specifies the effect of solution composition on the ratio of $\mathrm{Sr^{2+}}$ and $\mathrm{Ca^{2+}}$ kink sites (where $\mathrm{Sr^{2+}}$ and $\mathrm{Ca^{2+}}$ bind with the calcite crystal). In the Ion-by-Ion model, the relative preference for $\mathrm{Sr^{2+}}$ over $\mathrm{Ca^{2+}}$ being attached is held constant (like the D11 model, because the two cations attach onto the same type of kink site, $\mathrm{CO_3^{2-}}$, on the crystal surface). Meanwhile, the detachment preference, which can be described by the backward-reaction Sr/Ca partition coefficient, $K_\mathrm{b}$ (as discussed in detail in \ref{app:general}), varies with solution composition (unlike the D11 model, where the detachment preference is treated as constant). The description of this variation introduces new variables and adjustable parameters to the model, which in turn allows the Ion-by-Ion model to better account for correlation between $K$ and $\mathrm{\Delta^{44/40}Ca}$ compared to the D11 model \citep[e.g.,][]{jia2022model}. However, as we show in the following text, the Ion-by-Ion model still falls short in explaining the $\mathrm{\Delta^{88/86}Sr}$--$\mathrm{\Delta^{44/40}Ca}$ correlation. Under the Ion-by-Ion framework, the Sr isotope fractionation factor $\mathrm{\Delta^{88/86}Sr}$ can also be predicted without invoking immeasurable variables and parameters, when $\mathrm{\Delta^{44/40}Ca}$ and $K$ are both known, via the following relation (see \ref{app:ion-by-ion} for derivation),
\begin{linenomath} 
\begin{multline}
    \mathrm{\Delta^{88/86}Sr_\mathrm{IbI}} = \mathrm{\Delta^{88/86}Sr_{inf}} + \left( \mathrm{\Delta^{88/86}Sr_{eq}}-\mathrm{\Delta^{88/86}Sr_{inf}} \right) \\
    \times \left[  
    1 + \frac{K_\mathrm{M}}{K_\mathrm{inf}} \left(\frac{\mathrm{\Delta^{44/40}Ca_{M}}-\mathrm{\Delta^{44/40}Ca_{eq}}}{\mathrm{\Delta^{44/40}Ca_{eq}}-\mathrm{\Delta^{44/40}Ca_{inf}}} \right) \right],  \label{DELTA_Sr-DELTA_Ca-ibi}
\end{multline}
\end{linenomath} 
where $\mathrm{\Delta^{88/86}Sr_\mathrm{IbI}}$ is the Sr isotope fractionation factor predicted under the Ion-by-Ion model framework from the experimentally measured Ca isotope fractionation factor $\mathrm{\Delta^{44/40}Ca_{M}}$ and Sr/Ca partition coefficient $K_\mathrm{M}$. The prediction in $\mathrm{\Delta^{88/86}Sr}$ of the Ion-by-Ion model fits the data slightly better than that of the D11 model (Fig.~\ref{fig:previous_models}b), but the root-square mean error of $\mathrm{\Delta^{88/86}Sr}$ (in this case ${\sim}0.09\permil$) is still much larger than the analytical uncertainty \citep[${\sim}0.02\permil$;][]{bohm2012strontium, wang2023investigation}. Moreover, all data points fall on one side of the prediction, rather than lie randomly on both sides of it; this systemic misfit implies that the model does not adequately capture the system. As is the case for the D11 model, the discrepancy between observations and the Ion-by-Ion model predictions cannot be solved by adjusting the equilibrium Sr isotope fractionation factor $\mathrm{\Delta^{88/86}Sr_{eq}}$ (Fig.~\ref{fig:previous_models}c) or the kinetic-limit parameters, including $K_\mathrm{inf}$, $\mathrm{\Delta^{44/40}Ca_{inf}}$, and $\mathrm{\Delta^{88/86}Sr_{inf}}$ (see \ref{app:parameters} for details).

\section{A new model incorporating non-classical crystallization}  \label{sec:new}
A limitation of previous models, such as D11 and ion-by-ion, is their reliance on the assumption that crystallization is primarily governed by the classical mono-ion attachments \citep[e.g.,][]{depaolo2011surface, nielsen2012self, nielsen2013general, jia2022model}. Multiple lines of evidence have suggested that the forward reaction of calcite precipitation involves different crystallization mechanisms: While the attachment of single ions dominates at low-supersaturation (near-equilibrium) conditions,  direct attachment of large species, ranging from polymeric multi-ion complexes to fully formed nano-particles, becomes increasingly important at high-supersaturation (far-from-equilibrium) conditions \citep[e.g.,][]{de2015crystallization, ivanov2014oriented, putnis2021crystallization}. Although these multiple crystallization pathways contribute variously at different conditions, microscopic observations suggest that in most cases they occur concurrently \citep[e.g.][]{nielsen2014situ, de2015crystallization}. Here, we attempt to build a comprehensive model that integrates different crystallization mechanisms.

The fractionation of isotopes during the forward reaction is likely controlled by ion dehydration, as isotopes with higher dehydration frequency are preferentially precipitated onto crystal surface \citep[e.g.,][]{depaolo2011surface, hofmann2012ion}. However, this mechanism may not hold for trace element partitioning. Although $\mathrm{Sr^{2+}}$ is weakly hydrated and can be more frequently dehydrated than $\mathrm{Ca^{2+}}$, the Sr/Ca partition coefficient is observed to be smaller than one even in the fast-precipitating kinetic limit \citep[e.g.,][]{gabitov2006partitioning}, implying that $\mathrm{Sr^{2+}}$ is less preferred than $\mathrm{Ca^{2+}}$ by the forward reaction \citep{nielsen2013general}. The low Sr/Ca partition coefficient may be attributed to larger radius of $\mathrm{Sr^{2+}}$ (compared to $\mathrm{Ca^{2+}}$) and thus lower propensity to be incorporated into crystal \citep[e.g.,][]{zachara1991sorption}. Overall, the incorporation process plays a more important role in trace element partitioning.

The formation of multi-ion polymers or nano-particles likely also involves ion dehydration (also see Sec.~\ref{sec:diss}), which could induce a similar isotope fractionation effect to the ion-by-ion attachment process. Additionally, these multi-ion polymers and nano-particles exhibit less structural order and are more prone to strain and defects compared to flat crystal surfaces \citep[e.g.,][]{michel2010ordered}. Consequently, they are expected to incorporate a larger fraction of $\mathrm{Sr^{2+}}$ than the crystal surface. This phenomenon suggests that the attachment of polymers and/or nano-particles may enhance the incorporation of $\mathrm{Sr^{2+}}$ into the growing crystal, resulting in an elevated Sr/Ca partition coefficient for the forward reaction. Here, we assume that the classical and non-classical forward-reaction mechanisms have different levels of preference for $\mathrm{Sr^{2+}}$ over $\mathrm{Ca^{2+}}$. Thus, the variation in Sr/Ca partitioning during crystal precipitation depends largely on the competition between different forward reaction mechanisms. At low supersaturation when the system approaches equilibrium, the classical ion-by-ion attachment dominates. As the  supersaturation level increases, the contribution of polymers and/or nano-particles rises \citep[e.g.,][]{de2015crystallization}, making non-classical crystallization increasingly significant. At high supersaturation levels, the non-classical crystallization pathway is notably pronounced and likely dominant in the kinetic limit.

Mathematically, the inclusion of the non-classical crystallization introduces a new variable: the fractional contribution of non-classical crystallization mechanism, denoted as $f_\mathrm{N}$. This brings the total number of independent variables to three: (1) the overall forward-to-backward reaction rate ratio $R_\mathrm{f}/R_\mathrm{b}$, (2) the backward-reaction Sr/Ca partition coefficient $K_\mathrm{b}$, and (3) the fractional contribution of the non-classical crystallization mechanism $f_\mathrm{N}$ for the forward reaction. Experimental measurements provide three constraints, (1) the Ca isotope fractionation factor $\mathrm{\Delta^{44/40}Ca_{M}}$, (2) the Sr/Ca partition coefficient $K_\mathrm{M}$, and (3) the Sr isotope fractionation factor $\mathrm{\Delta^{88/86}Sr_{M}}$. Now, with an equal number of independent variables and constraints, for each experiment, we can find a unique solution for the variable system such that all constraints are together satisfied, or in another words, all observations can be explained. This makes our model distinct from previous models, which commonly have issues of ``overdetermination''. As an example, for the D11 model, the only variable $R_\mathrm{f}/R_\mathrm{b}$ can be conceivably determined by any one of $\mathrm{\Delta^{44/40}Ca_{M}}$, $K_\mathrm{M}$, and $\mathrm{\Delta^{88/86}Sr_{M}}$ alone, but no $R_\mathrm{f}/R_\mathrm{b}$ value could simultaneously satisfy all constraints of $\mathrm{\Delta^{44/40}Ca_{M}}$, $K_\mathrm{M}$, and $\mathrm{\Delta^{88/86}Sr_{M}}$ (as a result of which, the mutual correlations of these observables cannot be explained).

As illustrated above, the inclusion of non-classical crystallization into the kinetic model provides an opportunity to resolve the issue encountered by previous models. This outcome may seem counter-intuitive, considering that non-classical crystallization is assumed to induce a similar isotope fractionation effect to the classical pathway. The reason why our new model yields different predictions in Sr isotope fractionation can be understood as follows: As the new model explains the Sr/Ca partition coefficient differently from previous models, it necessitates a different presentation of the backward-reaction Sr/Ca partition coefficient $K_\mathrm{b}$. For example, the relation required by D11 model, $K_\mathrm{b} = K_\mathrm{inf}/K_\mathrm{eq}$ (see \ref{app:D11}), does not apply in our model. The Sr isotope fractionation factor during net precipitation is controlled by the ratio of $\mathrm{Sr^{2+}}$ attachment and $\mathrm{Sr^{2+}}$ detachment rates, and any change in the estimate of $K_\mathrm{b}$ would alter the estimate of the $\mathrm{Sr^{2+}}$ attachment/detachment ratio, and, hence, affects the prediction of $\mathrm{\Delta^{88/86}Sr}$ (the specific impact of $K_\mathrm{b}$ on $\mathrm{\Delta^{88/86}Sr}$ prediction is illustrated by Eq.~\ref{DELTA_88/86_final}). Consequently, our revised interpretation for the Sr/Ca partition coefficient leads to different predictions of $\mathrm{\Delta^{88/86}Sr}$ (and thus $\mathrm{\Delta^{88/86}Sr}$--$\mathrm{\Delta^{44/40}Ca}$ correlation), which potentially resolves the challenge described faced by previous models in light of the new Sr isotope observations (Sec.~\ref{sec:previous}).

At the same time, the measurements ($\mathrm{\Delta^{44/40}Ca_{M}}$, $\mathrm{\Delta^{88/86}Sr_{M}}$, and $K_\mathrm{M}$) provide enough constraints to uniquely solve the values of $R_\mathrm{f}/R_\mathrm{b}$, $K_\mathrm{b}$, and particularly the $f_\mathrm{N}$ at varying precipitation rates or solution supersaturation levels, permitting a unique application of this new model. The fractional contributions of the classical and non-classical crystallization mechanisms for the forward reaction, $f_\mathrm{C}$ and $f_\mathrm{N}$ ($f_\mathrm{C}$ + $f_\mathrm{N}$ = 1), can be estimated as (see \ref{app:non-classical} for detailed derivations)
\begin{linenomath} \begin{equation}
    f_\mathrm{C} = \frac{ K_\mathrm{inf} - K^\prime_\mathrm{M}}{ K_\mathrm{inf}-K^\prime_\mathrm{eq}}, \quad 
    f_\mathrm{N} = \frac{ K^\prime_\mathrm{M} - K^\prime_\mathrm{eq}}{ K_\mathrm{inf}-K^\prime_\mathrm{eq}},  \label{fCfN}
\end{equation}\end{linenomath} 
where $K^\prime_\mathrm{M}$ is an adjusted Sr/Ca partition coefficient calculated as
\begin{linenomath} \begin{equation}
    K^\prime_\mathrm{M} = K_\mathrm{M} \left( \frac{\mathrm{\Delta^{88/86}Sr_{eq}}-\mathrm{\Delta^{88/86}Sr_{M}}}{\mathrm{\Delta^{88/86}Sr_{eq}}-\mathrm{\Delta^{88/86}Sr_{inf}}} \right)^{-1} \left( \frac{\mathrm{\Delta^{44/40}Ca_{eq}}-\mathrm{\Delta^{44/40}Ca_{M}}}{\mathrm{\Delta^{44/40}Ca_{eq}}-\mathrm{\Delta^{44/40}Ca_{inf}}} \right),  \label{Kp_prime}
\end{equation}\end{linenomath} 
and $K^\prime_\mathrm{eq}$ is the adjusted equilibrium partition coefficient that is calculated as $K^\prime_\mathrm{eq} = K_\mathrm{eq} K_\mathrm{b(eq)}$, in which $K_\mathrm{b(eq)}$ is the equilibrium value of the backward-reaction partition coefficient, $K_\mathrm{b}$. The backward-reaction partition coefficient $K_\mathrm{b}$ (which describes the preference for Sr over Ca being detached from the crystal surface; also see \ref{app:general} for detailed discussion) can be estimated from isotope measurements as
\begin{linenomath} \begin{equation}
    K_\mathrm{b} = \left( \frac{\mathrm{\Delta^{88/86}Sr_{M}}-\mathrm{\Delta^{88/86}Sr_{inf}}}{\mathrm{\Delta^{88/86}Sr_{eq}}-\mathrm{\Delta^{88/86}Sr_{M}}} \right) \left( \frac{\mathrm{\Delta^{44/40}Ca_{M}}-\mathrm{\Delta^{44/40}Ca_{inf}}}{\mathrm{\Delta^{44/40}Ca_{eq}}-\mathrm{\Delta^{44/40}Ca_{M}}} \right)^{-1}.  \label{Eq:Kb}
\end{equation}\end{linenomath} 
At the equilibrium condition, $K_\mathrm{b}$ can no longer be constrained directly from Ca and Sr isotope measurements, because $\mathrm{\Delta^{44/40}Ca_{M}}$ and $\mathrm{\Delta^{88/86}Sr_{M}}$ would converge to their equilibrium values, $\mathrm{\Delta^{44/40}Ca_{eq}}$ and $\mathrm{\Delta^{88/86}Sr_{eq}}$, respectively. However, the value of $K_\mathrm{b(eq)}$ may be extrapolated from the results obtained from samples precipitated at relatively low supersaturation levels; in this study, we find $K_\mathrm{b(eq)} \approx 1$ with the experimental data for Ca and Sr isotope fractionations (see below for details). The overall forward reaction rate, $R_\mathrm{f}$, can be estimated using the net precipitation rate, $R_\mathrm{p}$, and Ca isotope fractionation factor as
\begin{linenomath} \begin{equation}
     R_\mathrm{f} = R_\mathrm{p} \times \left(\frac{ \mathrm{\Delta^{44/40}Ca_{inf}}-\mathrm{\Delta^{44/40}Ca_{eq}}}{\mathrm{\Delta^{44/40}Ca_{M}}-\mathrm{\Delta^{44/40}Ca_{eq}}}\right) .  \label{forward_rate}
\end{equation}\end{linenomath} 
The absolute forward-reaction rates associated with the classical and non-classical crystallization mechanisms can be calculated from the precipitation experiments as $R_\mathrm{f(C)} = f_\mathrm{C}R_\mathrm{f}$ and $R_\mathrm{f(N)} = f_\mathrm{N}R_\mathrm{f}$ using the above equations (Eqs.~\ref{fCfN}--\ref{forward_rate}).

With the Ca and Sr isotope measurements \citep{bohm2012strontium}, we find that the backward-reaction Sr/Ca partition coefficient $K_\mathrm{b}$ is approximately constant ($1.0 \pm 0.2$) under explored conditions (Fig.~\ref{fig:Kb}a). If we consider $K_\mathrm{b} = 1$ as a constant instead of a variable, then Eq.~\ref{Eq:Kb} is effectively a description of the $\mathrm{\Delta^{88/86}Sr}$--$\mathrm{\Delta^{44/40}Ca}$ correlation under our new model framework. In this case, we can use the measured $\mathrm{\Delta^{44/40}Ca}$ to predict $\mathrm{\Delta^{88/86}Sr}$ as
\begin{linenomath} 
\begin{equation}
    \mathrm{\Delta^{88/86}Sr_{NC}} = \mathrm{\Delta^{88/86}Sr_{inf}} + \left(\mathrm{\Delta^{88/86}Sr_{eq}}-\mathrm{\Delta^{88/86}Sr_{inf}} \right)
    \times \left( \frac{\mathrm{\Delta^{44/40}Ca_{M}}-\mathrm{\Delta^{44/40}Ca_{inf}}}{\mathrm{\Delta^{44/40}Ca_{eq}}-\mathrm{\Delta^{44/40}Ca_{inf}}} \right),  \label{DELTA_Sr_DELTA_Ca-linear}
\end{equation}
\end{linenomath} 
where $\mathrm{\Delta^{88/86}Sr_{NC}}$ is the Sr isotope fractionation factor predicted using the non-classical model (under the approximation of $K_\mathrm{b} = 1$) from the measured Ca isotope fractionation factor. In this new model, the predicted relationship of $\mathrm{\Delta^{88/86}Sr}$ and $\mathrm{\Delta^{44/40}Ca}$ corresponds closely to the experimental observations (Figs.~\ref{fig:Kb}b). The root-square mean error for $\mathrm{\Delta^{88/86}Sr}$ of this new prediction is ${\sim}0.02\permil$, which is better than any previous model and comparable to the analytical uncertainty \citep{bohm2012strontium, wang2023investigation}. With the approximation of $K_\mathrm{b} = 1$, we obtain from Eqs.~\ref{Kp_prime}--\ref{Eq:Kb} that $K^\prime_\mathrm{M} = K$ and $K^\prime_\mathrm{eq} = K_\mathrm{eq}$, and can simplify the fractional contributions of classical and non-classical mechanisms as
\begin{linenomath} \begin{equation}
    f_\mathrm{C} = \frac{K_\mathrm{inf}-K_\mathrm{M}}{K_\mathrm{inf}-K_\mathrm{eq}}, \quad
    f_\mathrm{N} = \frac{K_\mathrm{M}-K_\mathrm{eq}}{K_\mathrm{inf}-K_\mathrm{eq}}.  \label{fCfN_approx}
\end{equation}\end{linenomath} 
Our model suggests that the $\mathrm{\Delta^{88/86}Sr}$--$\mathrm{\Delta^{44/40}Ca}$ relation should be linear under the approximation of $K_\mathrm{b} = 1$ (Eq.~\ref{DELTA_Sr_DELTA_Ca-linear}). However, the same linear relation does not necessarily hold for $K$--$\mathrm{\Delta^{44/40}Ca}$. Eq.~\ref{fCfN_approx} suggests $K = K_\mathrm{eq} f_\mathrm{C} + K_\mathrm{inf} f_\mathrm{N}$, but the model itself does not predict how $f_\mathrm{C}$ and $f_\mathrm{N}$ vary with $\mathrm{\Delta^{44/40}Ca}$. Thus, the actual $K$--$\mathrm{\Delta^{44/40}Ca}$ relation should rely solely on observational evidence.

\section{Results}  \label{sec:res}
A notable contribution of the new model framework is its ability to quantitatively analyze and differentiate the contributions between the classical and non-classical crystallization mechanisms in calcite precipitation. Applying previously reported trace element and isotope measurements \citep{tang2008II, tang2008I,  bohm2012strontium}, we calculate the absolute forward reaction rates of the classical and non-classical crystallization pathways, and their fractional contributions at different conditions (Fig.~\ref{fig:reaction_rates}). While both crystallization pathways occur in most conditions, the relative contribution of the non-classical mechanism increases with the level of supersaturation. Specifically, the classical pathway dominates at low supersaturation, and non-classical pathway becomes more important than the classical mechanism at high supersaturation levels (where $\Omega = [\mathrm{Ca^{2+}}]_\mathrm{aq} [\mathrm{CO_3^{2-}}]_\mathrm{aq}/\mathcal{K}_\mathrm{sp} \gtrsim 15$, in which $\Omega$ is the saturation state, $[\mathrm{Ca^{2+}}]_\mathrm{aq}$ and $[\mathrm{CO_3^{2-}}]_\mathrm{aq}$ are the concentrations of $\mathrm{Ca^{2+}}$ and $\mathrm{CO_3^{2-}}$ in the aqueous solution, and $\mathcal{K}_\mathrm{sp}$ is the solubility product). Moreover, the new framework can be readily generalized to other isotope systems and trace elements. If a future study conducts joint isotope measurements on Ca, Sr, and other metal elements (such as Li or Ba), comparing predictions from different isotope systems could provide additional constraints to refine the theoretical framework presented here.

\section{Discussion}  \label{sec:diss}
In this work, we assume that the formation of multi-ion polymers or nano-particles involves dehydration of ions and, thus, is associated with the same isotope fractionation effect as the ion-by-ion attachment process. Previous studies suggest that the formation of weakly dehydrated amorphous calcium carbonate (ACC) particles leads to reduced isotope fractionation during the forward reaction \citep[e.g.,][]{lemarchand2004rate, alkhatib2017calcium, Lammers2021isotopic}. For this reason, the results of experiments involving ACC particles \citep{alkhatib2017calcium} are not considered in our calculations. However, due to the high solubility product of ACC compared to crystalline calcite, the effect of weakly dehydrated ACC is observed only in highly supersaturated solutions \citep[when the calcite saturation state, $\Omega$, exceeds ${\sim}20$; e.g.,][]{tang2008I, tang2008II, alkhatib2017calcium}. In growth experiments cnoducted with the calcite saturation states below ${\sim}20$ and a wide range of precipitation rate spanning nearly two orders of magnitude \citep{tang2008I, tang2008II}, the Ca isotope fractionation can be effectively accounted for by a constant forward-reaction Ca isotope fractionation \citep{depaolo2011surface}. Therefore, for these experiments, we disregard the effects of weakly dehydrated ACC and assume that the forward-reaction isotope fractionation factors remain constant. 

A limited number of studies have reported paired measurements of partition coefficients and isotope fractionation factors across multiple isotope systems. Given that the well-studied Ca and Sr systems yield a linear $\mathrm{\Delta^{88/86}Sr}$--$\mathrm{\Delta^{44/40}Ca}$ relation in both controlled experiments \citep{bohm2012strontium} and natural settings \citep{wang2021stable, shao2021impact, wang2023application}, it is reasonable to consider the possibility of a linear correlation between the fractionations of Ca and other metal isotope systems associated with calcite \citep{bohm2012strontium}. Nevertheless, our derivation here shows that this expectation may not hold true for all isotope systems. Indeed, the linear correlation between Ca and Sr isotope fractionation is not compatible with surface kinetic models within the classical model framework ({Figs.~\ref{fig:previous_models}, \ref{fig:D11M2}). Even within our model framework that incorporates the non-classical crystallization pathway, the existence of a linear correlation between Ca and Sr isotope fractionations implies that $K_\mathrm{b}$ (the backward-reaction Sr/Ca partition coefficient) is close to one for all conditions explored by previous experimental studies. While this approximation has been supported by experimental measurements for the Sr system (Fig.~\ref{fig:Kb}), it does not necessarily apply to other elements. It is possible that such an approximation holds for metallic cations that behave similarly to $\mathrm{Ca^{2+}}$ during carbonate dissolution, such as $\mathrm{Sr^{2+}}$ and $\mathrm{Ba^{2+}}$, but it does not hold for those that behave disparately from $\mathrm{Ca^{2+}}$, such as $\mathrm{Mg^{2+}}$, $\mathrm{Fe^{2+}}$, and $\mathrm{Mn^{2+}}$ \citep[e.g.,][]{pokrovsky2002surface}. Here, we suggest that any efforts to generalize such findings to other isotope systems that lack paired isotope measurements should be approached with caution. The current study emphasizes the importance of conducting simultaneous measurements of different isotope systems, which is crucial for improving the quantification of crystallization mechanisms and understanding the kinetic effects on trace element and isotope fractionation for their geochemical, geobiological, and environmental applications.

Previous studies have demonstrated that calcite crystals precipitated at different conditions exhibit different surface morphologies: At low supersaturation levels, the surfaces exhibit spiral structures; at high supersaturation levels, the surface structure is dominated by island-like two-dimensional nuclei; at intermediate supersaturation levels, the two structures coexist \citep[e.g.,][]{dove1993calcite, teng2000kinetics}. It is unclear how such morphological structures are related to the classical versus non-classical crystallization pathways. The classical models assume that both spiral and two-dimensional nuclei grow by mono-ion attachment, so that crystallization proceeds primarily by the classical pathway regardless of the dominant morphological structure \citep{nielsen2012self, nielsen2013general, jia2022model}. It is also possible that the transition in morphological structure is directly linked to the transition between classical and non-classical crystallization pathways. Multiple studies have reported the transition between spirals and two-dimensional nuclei in different saturation states \citep[e.g.,][]{dove1993calcite, gratz1993step, teng2000kinetics}. It is likely that, in addition to saturation state, the transition also relies on other conditions such as the solution pH and the calcium-to-carbonate ratio. Since the experiments were performed under different conditions, we cannot directly compare the morphological transitions reported by previous studies \citep[e.g.,][]{dove1993calcite, gratz1993step, teng2000kinetics} and the crystallization pathway transition suggested by this study. However, the framework presented here serves as a valuable tool for further studies aiming to establish the relationship between morphological structures and crystallization pathways. As the precipitation rate depends on both the morphological structure and the crystallization pathway \citep[e.g.,][]{van1993crystal, andersson2016microkinetic}, the application of our model framework would also provide critical insights for understanding crystal precipitation rates under different conditions. 

\section{Conclusion}
Modern microscopy observations have revealed that crystal growth occurs through both classical (ion-by-ion addition) and non-classical (polymer/particle attachment) crystallization pathways under most conditions. In this study, we present a new theoretical framework that incorporates both the classical and non-classical crystallization mechanisms to explicate the reaction kinetics during calcite precipitation from aqueous solution. This new development allows us to quantify the relative contributions of classical and non-classical crystallization under different conditions. We find that the relative contribution of the non-classical crystallization pathway increases with the level of supersaturation and becomes predominant at high supersaturation states. The new model readily explains the observed Sr partitioning and Ca and Sr isotope fractionation behaviors during calcite precipitation, and can be further extended to other mineral or isotope systems to rationalize their reaction kinetics. 

\section*{Acknowledgements}
The authors are grateful to L. Tarhan, N. Planavsky, X. Gu, and X. Chen for discussions, the reviewers for constructive reviews that improved this study, and the editor for handling the manuscript. J. Wang was supported by Peking University and the Agouron Geobiology Fellowship.

\clearpage
\begin{table}
\begin{center}
\begin{threeparttable}

\caption{Preferred equilibrium- and kinetic-limit parameters.} \label{Tab:parameters}

\begin{tabular}{p{4cm}p{3cm}p{3cm}p{3cm}}
    \hline 
    & $K$ & $\mathrm{\Delta^{44/40}Ca}$ & $\mathrm{\Delta^{88/86}Sr}$ \\
    \hline
    equilibrium limit & 0.025$^{1}$ & 0\permil$^{2}$ & 0\permil$^{3,4}$ \\
    \hline
    kinetic limit & 0.25$^{5,6}$ & -1.7\permil$^{4,6}$ & -0.34\permil$^{4}$ \\
    \hline
\end{tabular}

\begin{tablenotes}
    \small
    \item $^1$\cite{zhang2020equilibrium}; $^2$\cite{fantle2007isotopes}; $^3$\cite{bohm2012strontium}; $^4$see discussion in \ref{app:limit_parameters}; $^5$\cite{gabitov2006partitioning};  $^6$\cite{depaolo2011surface}.
\end{tablenotes}
    
\end{threeparttable}
\end{center}
\end{table}

\clearpage
\begin{figure}
\centering
    \includegraphics[width=1.0\textwidth]{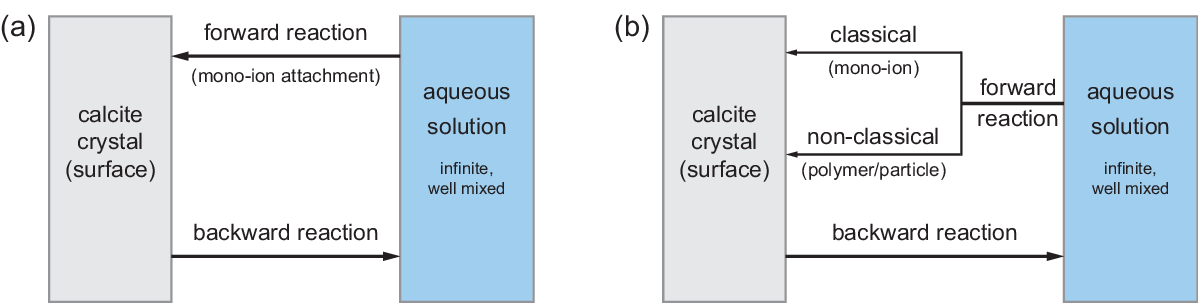}
    \caption{Schematic of the surface reaction models for calcite precipitation. (a) The classical view that the forward reaction occurs via mono-ion attachment only. (b) The current view that the forward reaction occurs via both classical (mono-ion attachment) and non-classical (multi-ion polymer or nano-particle attachment) crystallization mechanisms. The backward reaction (ion detachment) is considered identical for both models. }  \label{fig:cartoon}
\end{figure}

\clearpage
\begin{figure}
\centering
    \includegraphics[width=1.0\textwidth]{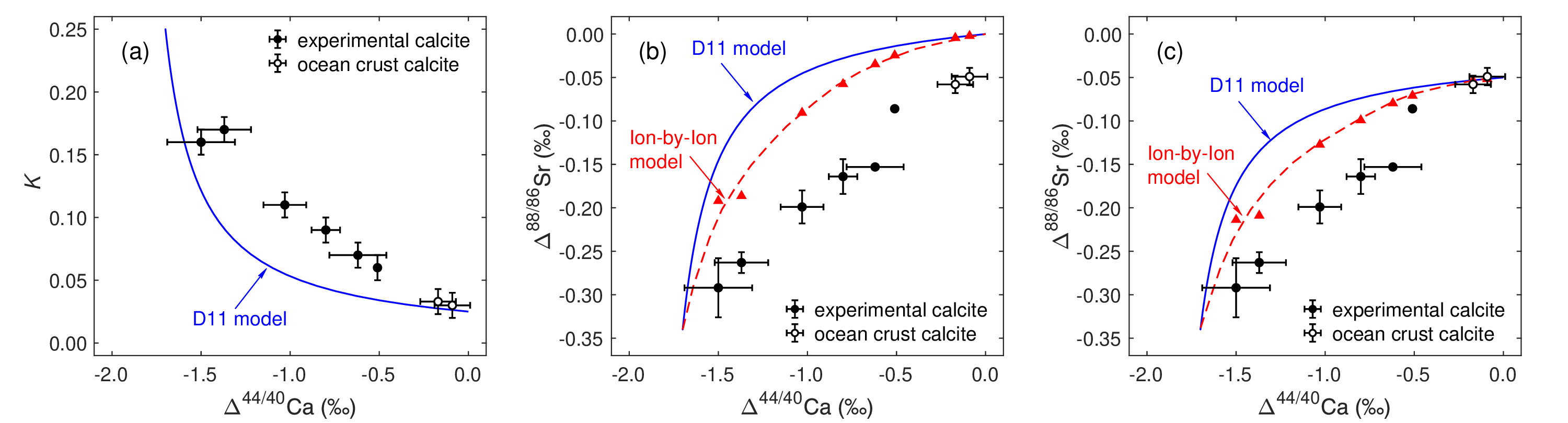}
    \caption{Observations of Sr/Ca partitioning and Ca and Sr isotope fractionation of experimental and ocean crust calcite \citep{bohm2012strontium} and comparison with predictions of previous models. (a) The correlation between Sr/Ca partition coefficient, $K$, and Ca isotope fractionation factor, $\mathrm{\Delta^{44/40}Ca}$. The blue curve is the D11 prediction (Eq.~\ref{KSr-DELTA_Ca-D11}). (b) The correlation between Sr isotope fractionation factor, $\mathrm{\Delta^{88/86}Sr}$, and Ca isotope fractionation factor, $\mathrm{\Delta^{44/40}Ca}$. The blue curve gives the D11 model prediction (Eq.~\ref{DELTA_Sr-DELTA_Ca-D11}). The red triangles represent the Ion-by-Ion model predictions (Eq.~\ref{DELTA_Sr-DELTA_Ca-ibi}). As the Ion-by-Ion model prediction for $\mathrm{\Delta^{88/86}Sr}$ involves the measured Sr/Ca partition coefficients (see Eq.~\ref{DELTA_Sr-DELTA_Ca-ibi}), we can only perform this calculation for the experimental data points for which the $K_\mathrm{M}$ values are available. However, these discrete points, together with the equilibrium- and kinetic-limit behaviors, allow us to draw an estimated continuous ``best fit'' (red dashed curve). The calculations are all performed with our preferred parameters for equilibrium and kinetic limits (Tab.~\ref{Tab:parameters}). (c) Same as (b) but with $\mathrm{\Delta^{88/86}Sr}_\mathrm{eq} = -0.05\permil$ for calculations, as it represents the only actual observation from the slowly precipitated calcite \citep[]{bohm2012strontium}.} \label{fig:previous_models}

\end{figure} 

\clearpage
\begin{figure}
\centering
    \includegraphics[width=1.0\textwidth]{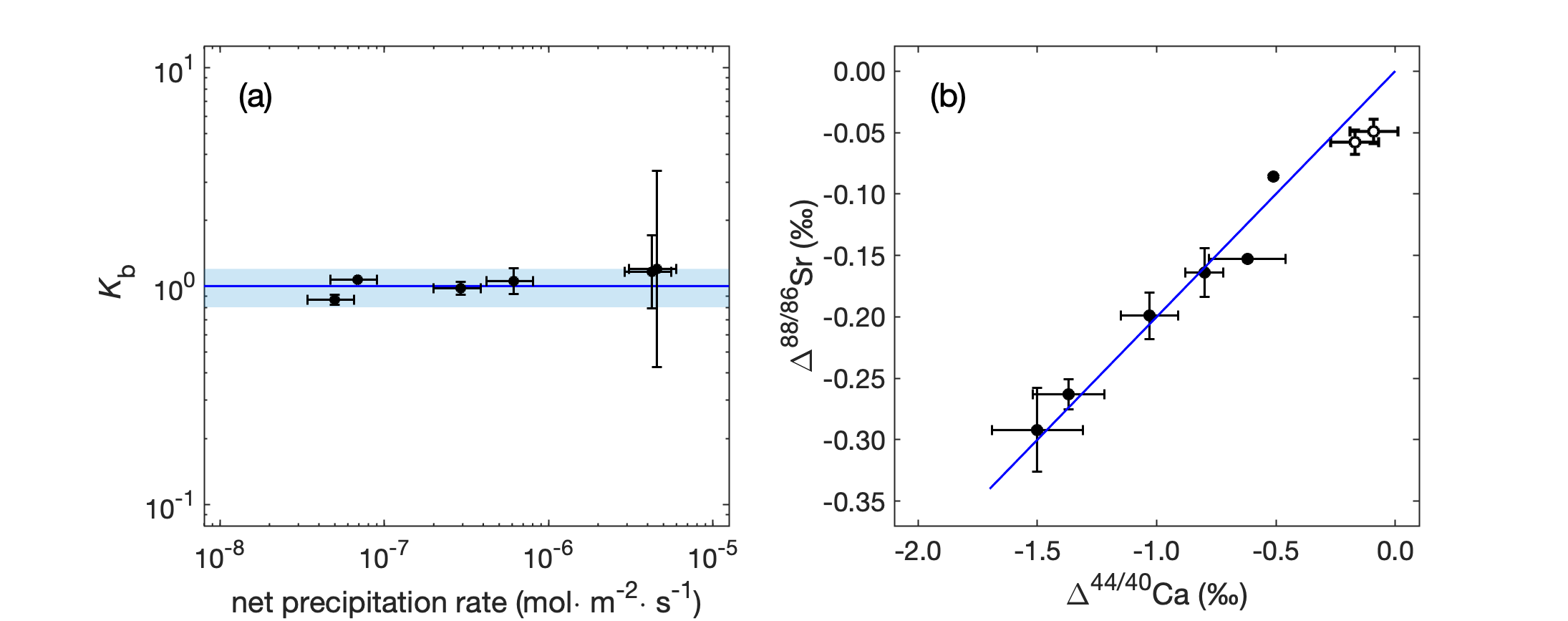}
    \caption{(a) The backward-reaction Sr/Ca partition coefficients estimated from the Ca and Sr isotope fractionation factors at different precipitation rates (using Eq.~\ref{Eq:Kb}). The Ca and Sr isotope fractionation factors and precipitation rates are all obtained directly from experiments \citep{bohm2012strontium}. The uncertainties in $K_\mathrm{b}$ are calculated from those in $\mathrm{\Delta^{44/40}Ca}$ and $\mathrm{\Delta^{88/86}Sr}$ measurements with error propagation. The blue line indicates $K_\mathrm{b} = 1$, and the light blue area indicates the range of $1 \pm 0.2$. (b) The comparison of experimental and natural observations (solid and open black circles) with the predicted $\mathrm{\Delta^{44/40}Ca}$--$\mathrm{\Delta^{88/86}Sr}$ correlation (blue curve) based on the proposed model framework with the approximation of $K_\mathrm{b} = 1$ (i.e., using Eq.~\ref{DELTA_Sr_DELTA_Ca-linear}).}   \label{fig:Kb}
\end{figure} 

\clearpage
\begin{figure}
\centering
    \includegraphics[width=1.0\textwidth]{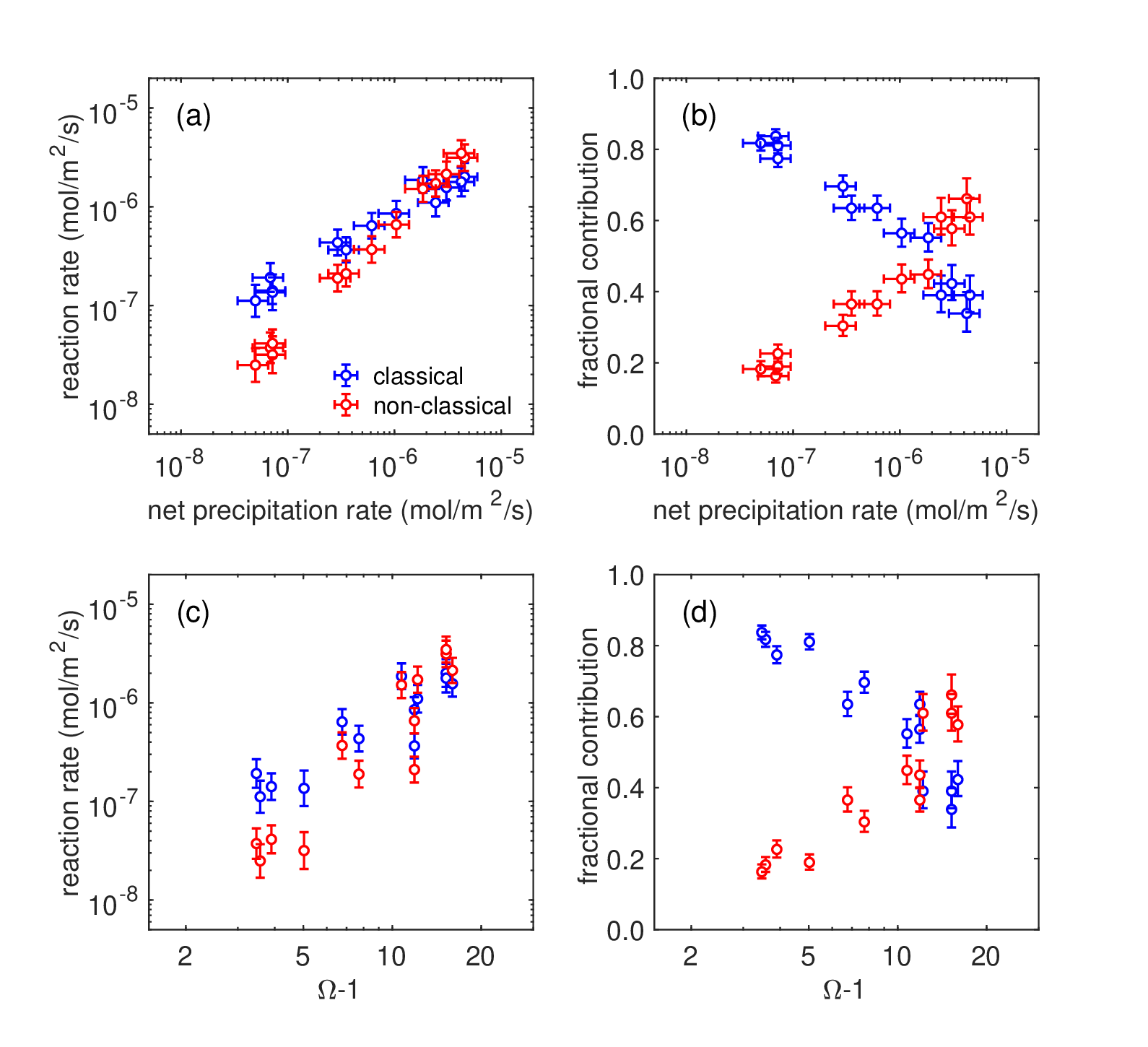}
    \caption{The absolute forward reaction rates (a, c) and the fractional contributions (b, d) of the classical crystallization mechanism (mono-ion attachment; blue) and the non-classical crystallization mechanism (multi-ion polymer or particle attachment; red) at different net crystal precipitation rates (a, b) or calcite saturation states $\Omega$ (c, d). All calculations are performed using the presented model (Eq.~\ref{fCfN_approx}) and measurements from experiments at $25\ \mathrm{^{\circ}C}$ \citep{tang2008I, tang2008II, bohm2012strontium}.} \label{fig:reaction_rates}
\end{figure}

\clearpage
\appendix
\section{Equilibrium and kinetic limits}  \label{app:limit_parameters}
The equilibrium parameters ($K_\mathrm{eq}$, $\mathrm{\Delta^{44/40}Ca_{eq}}$, and $\mathrm{\Delta^{88/86}Sr_{eq}}$) could be constrained by direct measurement of fluids and crystals precipitated at chemical and isotope equilibrium. Analyses of extremely slowly precipitated sediments ($R_\mathrm{p}$ = $10^{-18}$--$10^{-17}\ \mathrm{mol\cdot{m}^{-2}\cdot{s}^{-1}}$) and pore fluids demonstrate that $K_\mathrm{eq} = 0.025 \pm 0.005$ at $25\ \mathrm{^\circ{C}}$ \citep{zhang2020equilibrium} and $\mathrm{\Delta^{44/40}Ca_{eq}} = 0.0 \pm 0.1 \permil$ \citep{fantle2007isotopes, jacobson2008delta44ca}. This $\mathrm{\Delta^{44/40}Ca_{eq}}$ value has recently been confirmed by controlled experiments \citep[][]{harrison2023equilibrium}. The value of $\mathrm{\Delta^{88/86}Sr_{eq}}$ is less well constrained, but the low Sr isotope fractionation in ocean crust calcite samples implies that the fractionation of Sr isotopes at equilibrium is weak, with a magnitude of less than $0.05\permil$ \citep{bohm2012strontium}. \cite{depaolo2011surface} suggested that $\mathrm{\Delta^{44/40}Ca_{eq}}$ is approximately $\mathrm{0\permil}$ because the preference of light isotopes during ion dehydration is compensated by the equal preference of light isotopes during ion re-hydration at equilibrium. Adopting this explanation, we may also expect an equilibrium fractionation of $0\permil$ for isotopes of other elements. This is supported by the experimental data inferring that the equilibrium fractionation of $\mathrm{^{137}Ba}$ and $\mathrm{^{134}Ba}$ between calcite and aqueous solutions is zero within uncertainty \cite[][]{mavromatis2020experimental}. Therefore, we assume $\mathrm{\Delta^{44/40}Ca_{eq}} = 0 \permil$, $\mathrm{\Delta^{88/86}Sr_{eq}} = 0 \permil$, and $K_\mathrm{eq} = 0.025$ as the preferred values throughout this manuscript.

The parameters for the kinetic limit can be inferred from experimental and natural observations. The value of $\mathrm{\Delta^{44/40}Ca_{inf}}$ corresponds to the most negative $\mathrm{\Delta^{44/40}Ca}$ that can be reached during calcite precipitation. The observed value of $\mathrm{\Delta^{44/40}Ca}$ ranges from $-0.1\permil$ to $-1.6\permil$ under different laboratory conditions \cite[e.g.,][]{lemarchand2004rate, tang2008II, mills2021influence} with an average of approximately $-1.4\permil$ in natural settings \cite[e.g.,][]{de2000isotopic,depaolo2004calcium, fantle2005variations, gussone2005calcium, gussone2009critical, kisakurek2011controls, gussone2016calcium}. In \cite{depaolo2011surface}, $\mathrm{\Delta^{44/40}Ca_{inf}} = -1.7 \permil$ was used to fit the experiments of \cite{tang2008II}. For Sr isotope fractionation, \cite{bohm2012strontium} and \cite{alkhatib2017calcium} found that $\mathrm{\Delta^{88/86}Sr}$ ranges between $-0.3\permil$ and $-0.1\permil$ under different precipitation rates in synthetic experiments, similar to the range observed in natural settings \cite[e.g.,][]{fietzke2006determination, krabbenhoft2010constraining, stevenson2014controls, voigt2015variability}. Their analysis also suggested that $\mathrm{\Delta^{88/86}Sr} \approx 0.2 \times \mathrm{\Delta^{44/40}Ca}$ at high precipitation rates, which is consistent with the results of molecular dynamics simulations on ion desolvation processes that suggest $\mathrm{\Delta^{88/86}Sr_{inf}} \approx 0.2 \times \mathrm{\Delta^{44/40}Ca_{inf}}$ \citep{hofmann2012ion}. Thus, we assume that $\mathrm{\Delta^{88/86}Sr_{inf}}=-0.34\permil$. Controlled laboratory studies suggest that the Sr/Ca partition coefficient increases with precipitation rate but flattens under the high-precipitation-rate limit with values scattered between ${\sim}0.25$ and ${\sim}0.35$ \citep{lorens1981sr, tesoriero1996solid, gabitov2006partitioning, gabitov2014crystal, tang2008I}. Therefore, we suggest the following parameters: $\mathrm{\Delta^{44/40}Ca_{inf}} = -1.7 \permil$, $\mathrm{\Delta^{88/86}Sr_{inf}}=-0.34\permil$, and $K_\mathrm{inf} = 0.25$.

\section{Surface kinetic models} \label{app:general}

\subsection{General framework}  
The net precipitation of calcite crystals from aqueous solution occurs as the overall result of the forward reaction (ion or particle attachment onto the crystal surface) and the backward reaction (ion detachment from the crystal surface). The Ca isotope fractionation during the forward and backward reactions is quantified using the fractionation factors defined as
\begin{linenomath} \begin{align}
    {^{44/40}\alpha}_\mathrm{f} = \frac{{^{44}R}_\mathrm{f}/{^{40}R}_\mathrm{f}} {(\mathrm{^{44}Ca}/\mathrm{^{40}Ca})_\mathrm{aq}},  \label{af_44/40} \\
    {^{44/40}\alpha}_\mathrm{b} = \frac{{^{44}R}_\mathrm{b}/{^{40}}R_\mathrm{b}} {(\mathrm{^{44}Ca}/\mathrm{^{40}Ca})_\mathrm{cal}},  \label{ab_44/40}
\end{align}\end{linenomath}
where ${^{40}R}_\mathrm{f}$ and ${^{44}R}_\mathrm{f}$ (${^{40}R}_\mathrm{b}$ and ${^{44}R}_\mathrm{b}$) are the forward (backward) reaction rates of $\mathrm{^{40}Ca}$ and $\mathrm{^{44}Ca}$, and $(\mathrm{^{44}Ca}/\mathrm{^{40}Ca})_\mathrm{aq}$ and $(\mathrm{^{44}Ca}/\mathrm{^{40}Ca})_\mathrm{cal}$ are the Ca isotope ratios of the aqueous solution and the calcite surface, respectively. The Ca isotope fractionation during net calcite precipitation from aqueous solutions is quantified by the fractionation factor
\begin{linenomath} \begin{equation}
    {^{44/40}\alpha} = \frac{{^{44}R}_\mathrm{p}/{^{40}R}_\mathrm{p}} {(\mathrm{^{44}Ca}/\mathrm{^{40}Ca})_\mathrm{aq}},  \label{ap44/40}
\end{equation}\end{linenomath} 
where ${^{40}R}_\mathrm{p} = {^{40}R}_\mathrm{f}-{^{40}R}_\mathrm{b}$ and ${^{44}R}_\mathrm{p} = {^{44}R}_\mathrm{f}-{^{44}}R_\mathrm{b}$ are the net precipitation rates of $\mathrm{^{40}Ca}$ and $\mathrm{^{44}Ca}$. In the steady state, the composition of the crystal remains constant, and the net precipitation of $\mathrm{^{44}Ca}$ and $\mathrm{^{40}Ca}$ in an infinitesimal time interval follows the $\mathrm{^{44}Ca/^{40}Ca}$ ratio in the existing crystal, such that
\begin{linenomath} \begin{equation}
    {^{44}R}_\mathrm{p}/{^{40}R}_\mathrm{p} = (\mathrm{^{44}Ca}/\mathrm{^{40}Ca})_\mathrm{cal}.  \label{steadystate}
\end{equation}\end{linenomath} 
Substituting Eqs.~\ref{af_44/40}, \ref{ab_44/40} and \ref{steadystate} into Eq.~\ref{ap44/40}, we obtain the following expression for the Ca isotope fractionation factor during calcite precipitation,
\begin{linenomath} \begin{equation}
    {^{44/40}\alpha} = \frac{ {^{44/40}\alpha}_\mathrm{f} (^{40}R_\mathrm{f}/^{40}R_\mathrm{b})}{ ^{40}R_\mathrm{f}/^{40}R_\mathrm{b} + ({^{44/40}\alpha}_\mathrm{b}-1)},  \label{ap_44/40_afab}
\end{equation}\end{linenomath} 
If ${^{44/40}\alpha}_\mathrm{f}$ and ${^{44/40}\alpha}_\mathrm{b}$ are both constants, the above expression can be shown to be equivalent to Eq.~11 of \cite{depaolo2011surface}. Applying the $\Delta$-notation for isotope fractionation, we obtain
\begin{linenomath} \begin{equation}
    \mathrm{\Delta^{44/40}Ca} = 1000\permil \times ({^{44/40}\alpha}_\mathrm{f}-1) - \frac{1000\permil \times ({^{44/40}\alpha}_\mathrm{f}{^{44/40}\alpha}_\mathrm{b} - {^{44/40}\alpha}_\mathrm{f})}{ ^{40}R_\mathrm{f}/^{40}R_\mathrm{b} + ({^{44/40}\alpha}_\mathrm{b}-1)}.  \label{DELTA_44/40_afab}
\end{equation}\end{linenomath} 

Considering the Sr/Ca partition processes, we define the $\mathrm{^{88}Sr}$/$\mathrm{^{40}Ca}$ partition coefficients during the forward and backward reactions,
\begin{linenomath} \begin{align}
    {^{88/40}K}_\mathrm{f} = \frac{^{88}R_\mathrm{f}/^{40}R_\mathrm{f}} {(\mathrm{^{88}Sr}/\mathrm{^{40}Ca})_\mathrm{aq}},  \\
    {^{88/40}K}_\mathrm{b} = \frac{^{88}R_\mathrm{b}/^{40}R_\mathrm{b}} {(\mathrm{^{88}Sr}/\mathrm{^{40}Ca})_\mathrm{cal}}.  \label{Kf_Kb_88/40}
\end{align}\end{linenomath}
In the D11 \citep{depaolo2011surface} model, ${^{88/40}K}_\mathrm{f}$ and ${^{88/40}K}_\mathrm{b}$ are both considered constants. In the Ion-by-Ion model \cite[e.g.,][]{nielsen2013general, jia2022model}, ${^{88/40}K}_\mathrm{f}$ is constant, while ${^{88/40}K}_\mathrm{b}$ varies with the solution chemistry (pH, saturation state, and $\mathrm{Ca^{2+}}:\mathrm{CO_3^{2+}}$ ratio). Whichever the case, if we apply the definitions given in Eq.~\ref{Kf_Kb_88/40}, we can follow the same derivation as that of Eq.~\ref{ap_44/40_afab} and express the $\mathrm{^{88}Sr}$/$\mathrm{^{40}Ca}$ partition coefficient during net precipitation as
\begin{linenomath} \begin{equation}
    {^{88/40}K} = \frac{^{88}R_\mathrm{p}/^{40}R_\mathrm{p}} {(\mathrm{^{88}Sr}/\mathrm{^{40}Ca})_\mathrm{aq}} = \frac{ {^{88/40}K}_\mathrm{f} ({^{40}R}_\mathrm{f}/{^{40}R}_\mathrm{b})}{ {^{40}R}_\mathrm{f}/{^{40}R}_\mathrm{b} + ({^{88/40}K}_\mathrm{b}-1) }.  \label{Kp_88/40}
\end{equation}\end{linenomath} 
Now we consider the Sr isotope fractionation. Modifying Eq.~\ref{Kp_88/40} by substituting $\mathrm{^{86}Sr}$ for $\mathrm{^{88}Sr}$, we express the $\mathrm{^{86}Sr}$/$\mathrm{^{40}Ca}$ partition coefficient during net precipitation as
\begin{linenomath} \begin{equation}
    {^{86/40}K} = \frac{{^{86}R}_\mathrm{p}/{^{40}R}_\mathrm{p}} {(\mathrm{^{86}Sr}/\mathrm{^{40}Ca})_\mathrm{aq}} = \frac{ {^{86/40}K}_\mathrm{f} ({^{40}R}_\mathrm{f}/{^{40}R}_\mathrm{b})}{ {^{40}R}_\mathrm{f}/{^{40}R}_\mathrm{b} + ({^{86/40}K}_\mathrm{b}-1) }.  \label{Kp_86/40}
\end{equation}\end{linenomath} 
Applying Eqs.~\ref{Kp_88/40} and \ref{Kp_86/40}, we express the Sr isotope fractionation factor during net precipitation, ${^{88/86}\alpha} = (^{88}R_\mathrm{p}/^{86}R_\mathrm{p})/(\mathrm{^{88}Sr}/\mathrm{^{86}Sr})_\mathrm{aq}$, as
\begin{linenomath} \begin{equation}
    {^{88/86}\alpha} = \frac{{^{88/40}K}}{{^{86/40}K}} = \frac{ {^{88/86}\alpha}_\mathrm{f} ({^{40}R}_\mathrm{f}/{^{40}R}_\mathrm{b} + {^{88/40}K}_\mathrm{b}/{^{88/86}\alpha}_\mathrm{b} - 1)}{ {^{40}R}_\mathrm{f}/{^{40}R}_\mathrm{b} + ({^{88/40}K}_\mathrm{b}-1)},  \label{ap_88/86_afab}
\end{equation}\end{linenomath} 
in which ${^{88/86}\alpha}_\mathrm{f}$ and ${^{88/86}\alpha}_\mathrm{b}$ are the Sr isotope fractionation factors during the forward and backward reactions that are defined respectively as
\begin{linenomath} \begin{align}
    {^{88/86}\alpha}_\mathrm{f} = \frac{{^{88}R}_\mathrm{f}/{^{86}R}_\mathrm{f}} {(\mathrm{^{88}Sr}/\mathrm{^{86}Sr})_\mathrm{aq}} = \frac{{^{88/40}K}_\mathrm{f}}{{^{86/40}K}_\mathrm{f}},  \\
    {^{88/86}\alpha}_\mathrm{b} = \frac{{^{88}R}_\mathrm{b}/{^{86}R}_\mathrm{b}} {(\mathrm{^{88}Sr}/\mathrm{^{86}Sr})_\mathrm{cal}} = \frac{{^{88/40}K}_\mathrm{b}}{{^{86/40}K}_\mathrm{b}}.  \label{af_ab_88/86}
\end{align}\end{linenomath}
Applying the $\Delta$-notation, we obtain
\begin{linenomath} \begin{equation}
    \mathrm{\Delta^{88/86}Sr} = 1000\permil \times ({^{88/86}\alpha}_\mathrm{f}-1) - 1000\permil \times\frac{ ({^{88/86}\alpha}_\mathrm{f}-{^{88/86}\alpha}_\mathrm{f}/^{88/86}\alpha_\mathrm{b}){^{88/40}K}_\mathrm{b}}{ ^{40}R_\mathrm{f}/^{40}R_\mathrm{b} + ({^{88/40}K}_\mathrm{b}-1)}.  \label{DELTA_88/86_afab}
\end{equation}\end{linenomath} 

\subsection{Classical models}  \label{app:classical}
In the D11 \citep{depaolo2011surface} and ion-by-ion \citep{nielsen2012self} models for Ca isotope fractionation, the forward- and backward-reaction Ca isotope fractionation factors ($^{44/40}\alpha_\mathrm{f}$ and $^{44/40}\alpha_\mathrm{b}$) are constants. If we extend these models to Sr isotope fractionation, the forward- and backward-reaction Sr isotope fractionation factors ($^{88/86}\alpha_\mathrm{f}$ and $^{88/86}\alpha_\mathrm{b}$) should also be considered constant. The fractionations of Ca and Sr isotopes (Eqs.~\ref{DELTA_44/40_afab} and \ref{DELTA_88/86_afab}) in the equilibrium ($^{40}R_\mathrm{f}/^{40}R_\mathrm{b} \to 0$) and kinetic ($^{40}R_\mathrm{f}/^{40}R_\mathrm{b} \to \infty$) limits lead to
\begin{linenomath} \begin{equation}
    \mathrm{\Delta^{44/40}Ca_{eq}} = 1000\permil \times ({^{44/40}\alpha}_\mathrm{f}/{^{44/40}\alpha}_\mathrm{b}-1), \label{44/40_eq}
\end{equation}
\begin{equation}
    \mathrm{\Delta^{44/40}Ca_{inf}} = 1000\permil \times ({^{44/40}\alpha}_\mathrm{f}-1), \label{44/40_inf}  
\end{equation}
\begin{equation}
    \mathrm{\Delta^{88/86}Sr_{eq}} = 1000\permil \times ({^{88/86}\alpha}_\mathrm{f}/{^{88/86}\alpha}_\mathrm{b}-1), \label{88/86_eq}  
\end{equation}
\begin{equation}
    \mathrm{\Delta^{88/86}Sr_{inf}} = 1000\permil \times ({^{88/86}\alpha}_\mathrm{f}-1),  \label{88/86_inf}
\end{equation}\end{linenomath} 
where $\mathrm{\Delta^{44/40}Ca_{eq}}$ and $\mathrm{\Delta^{88/86}Sr_{eq}}$ are Ca and Sr isotope fractionation in the equilibrium limit, and $\mathrm{\Delta^{44/40}Ca_{inf}}$ and $\mathrm{\Delta^{88/86}Sr_{inf}}$ are Ca and Sr isotope fractionation in the kinetic limit. From Eqs.~\ref{44/40_eq} and \ref{44/40_inf}, we obtain the expressions of ${^{44/40}\alpha}_\mathrm{f}$ and ${^{44/40}\alpha}_\mathrm{b}$,
\begin{linenomath} \begin{equation}
    {^{44/40}\alpha}_\mathrm{f} = 1 + \frac{\mathrm{\Delta^{44/40}Ca_{inf}}}{1000\permil},  \label{44/40_alpha_f} 
\end{equation}
\begin{equation}
    {^{44/40}\alpha}_\mathrm{b} = \frac{1000\permil + \mathrm{\Delta^{44/40}Ca_{inf}}}{ 1000\permil + \mathrm{\Delta^{44/40}Ca_{eq}} } = 1 + \epsilon,  \label{44/40_alpha_eq}
\end{equation}\end{linenomath} 
where $\epsilon = {^{44/40}\alpha}_\mathrm{b} - 1$ is expressed as
\begin{linenomath} \begin{equation}
    \epsilon = \frac{\mathrm{\Delta^{44/40}Ca_{inf}} - \mathrm{\Delta^{44/40}Ca_{eq}}}{1000\permil + \mathrm{\Delta^{44/40}Ca_{eq}}}.  \label{epsilon}
\end{equation}\end{linenomath} 
Substituting Eqs.~\ref{44/40_alpha_f} and \ref{44/40_alpha_eq} into Eq.~\ref{DELTA_44/40_afab}, we obtain
\begin{linenomath} \begin{equation}
    \mathrm{\Delta^{44/40}Ca} = \mathrm{\Delta^{44/40}Ca_{inf}} - \frac{ \left( 1000\permil + \mathrm{\Delta^{44/40}Ca_{inf}} \right) \epsilon }{  {^{40}R}_\mathrm{f}/{^{40}R}_\mathrm{b} + \epsilon}.  \label{DELTA_44/40_Ca_add_review_01}
\end{equation}\end{linenomath} 
With the expression of $\epsilon$ given by Eq.~\ref{epsilon}, we obtain $(1000\permil + \mathrm{\Delta^{44/40}Ca_{inf}})\epsilon = (\mathrm{\Delta^{44/40}Ca_{inf}} - \mathrm{\Delta^{44/40}Ca_{eq}}) [(1000\permil + \mathrm{\Delta^{44/40}Ca_{inf}})/(1000\permil + \mathrm{\Delta^{44/40}Ca_{eq}})] = (\mathrm{\Delta^{44/40}Ca_{inf}} - \mathrm{\Delta^{44/40}Ca_{eq}})(1+\epsilon)$. Therefore, we can write Eq.~\ref{DELTA_44/40_Ca_add_review_01} as
\begin{linenomath} 
\begin{equation}
    \mathrm{\Delta^{44/40}Ca} = \mathrm{\Delta^{44/40}Ca_{inf}} + \frac{ (\mathrm{\Delta^{44/40}Ca_{eq}} - \mathrm{\Delta^{44/40}Ca_{inf}}) (1+\epsilon) }{  {^{40}R}_\mathrm{f}/{^{40}R}_\mathrm{b} + \epsilon}.
\end{equation}
\end{linenomath} 
Since $\epsilon \ll 1 <{^{40 }R}_\mathrm{f}/{^{40}R}_\mathrm{b}$ during net calcite precipitation, the contributions of $\epsilon$ would be negligibly small; thus, we can write the above equation approximately as
\begin{linenomath} 
\begin{equation}
    \mathrm{\Delta^{44/40}Ca} = \mathrm{\Delta^{44/40}Ca_{inf}} + \frac{ \mathrm{\Delta^{44/40}Ca_{eq}} - \mathrm{\Delta^{44/40}Ca_{inf}} }{  {^{40}R}_\mathrm{f}/{^{40}R}_\mathrm{b}}.  \label{DELTA_44/40_final}
\end{equation}
\end{linenomath}
Taking the difference between Eqs.~\ref{88/86_eq} and \ref{88/86_inf} yields
\begin{linenomath} \begin{equation}
     1000\permil \times ({^{88/86}\alpha}_\mathrm{f}-{^{88/86}\alpha}_\mathrm{f}/{^{88/86}\alpha}_\mathrm{b}) =     \mathrm{\Delta^{88/86}Sr_{inf}} -\mathrm{\Delta^{88/86}Sr_{eq}},  \label{DELTA_88/86_Sr_add_review_01}
\end{equation}\end{linenomath} 
Substituting Eq.~\ref{DELTA_88/86_Sr_add_review_01} into Eq.~\ref{DELTA_88/86_afab}, we obtain
\begin{linenomath} 
\begin{equation}
    \mathrm{\Delta^{88/86}Sr} = \mathrm{\Delta^{88/86}Sr_{inf}} + \frac{ ( \mathrm{\Delta^{88/86}Sr_{eq}}-\mathrm{\Delta^{88/86}Sr_{inf}}) {^{88/40}K}_\mathrm{b}}{ ^{40}R_\mathrm{f}/^{40}R_\mathrm{b} + ({^{88/40}K}_\mathrm{b}-1)}. \label{DELTA_88/86_final}
\end{equation}
\end{linenomath} 
Eqs.~\ref{Kp_88/40}, \ref{DELTA_44/40_final}, and \ref{DELTA_88/86_final} are the general expressions of Sr/Ca partition, Ca isotope fractionation, and Sr isotope fractionation, under the framework of surface kinetic model and the assumption of constant forward- and backward-reaction isotope fractionation factors.

\subsubsection{D11 model}  \label{app:D11}
In the D11 \citep{depaolo2011surface} model, ${^{88/40}K}_\mathrm{f}$ and ${^{88/40}K}_\mathrm{b}$ are both constant. In this case, considering the $\mathrm{^{88}Sr}/\mathrm{^{40}Ca}$ partitioning (Eq.~\ref{Kp_88/40}) in the limits of equilibrium ($^{40}R_\mathrm{f}/^{40}R_\mathrm{b} \to 0$) and far-from-equilibrium ($^{40}R_\mathrm{f}/^{40}R_\mathrm{b} \to \infty$) conditions, we obtain
\begin{linenomath} \begin{equation}
    {^{88/40}K}_\mathrm{eq} = {^{88/40}K}_\mathrm{f}/{^{88/40}K}_\mathrm{b},  \label{Keq_D11} 
\end{equation}
\begin{equation}
    {^{88/40}K}_\mathrm{inf} = {^{88/40}K}_\mathrm{f},  \label{Kf_D11}
\end{equation}\end{linenomath} 
where ${^{88/40}K}_\mathrm{eq}$ and ${^{88/40}K}_\mathrm{inf}$ are $\mathrm{^{88}Sr}/\mathrm{^{40}Ca}$ partition coefficients in the equilibrium limit and the kinetic limit. Substituting these relations into Eq.~\ref{Kp_88/40} and Eq.~\ref{DELTA_88/86_final} leads to 
\begin{linenomath} 
\begin{equation}
    {^{88/40}K} = {^{88/40}K}_\mathrm{inf} +  
    \frac{({^{88/40}K}_\mathrm{eq}-{^{88/40}K}_\mathrm{inf})({^{88/40}K}_\mathrm{inf}/{^{88/40}K}_\mathrm{eq})}{ ^{40}R_\mathrm{f}/^{40}R_\mathrm{b} + ({^{88/40}K}_\mathrm{inf}/{^{88/40}K}_\mathrm{eq}-1)},  \label{Kp_88/40_D11} 
\end{equation}
\begin{equation}
    \mathrm{\Delta^{88/86}Sr} = \mathrm{\Delta^{88/86}Sr_{inf}} + 
    \frac{(\mathrm{\Delta^{88/86}Sr_{eq}}-\mathrm{\Delta^{88/86}Sr_{inf}}) ({^{88/40}K}_\mathrm{inf}/{^{88/40}K}_\mathrm{eq})}{ ^{40}R_\mathrm{f}/^{40}R_\mathrm{b} + ({^{88/40}K}_\mathrm{inf}/{^{88/40}K}_\mathrm{eq}-1)}.  \label{DELTA88/86_D11}
\end{equation}
\end{linenomath} 
To test the model against the experimental data of $K$--$\mathrm{\Delta^{44/40}Ca}$ and $\mathrm{\Delta^{88/86}Sr}$--$\mathrm{\Delta^{44/40}Ca}$ correlations, we combine Eqs.~\ref{DELTA_44/40_final}, \ref{Kp_88/40_D11}, and \ref{DELTA88/86_D11} together, and after some algebra, obtain
\begin{linenomath} 
\begin{multline}
    {^{88/40}K} = {^{88/40}K}_\mathrm{inf} + \left({^{88/40}K}_\mathrm{eq}-{^{88/40}K}_\mathrm{inf}\right) \\
    \times \left(\frac{{^{88/40}K}_\mathrm{inf}}{{^{88/40}K}_\mathrm{eq}}\right) \left(  \frac{ \mathrm{\Delta^{44/40}Ca_{eq}}-\mathrm{\Delta^{44/40}Ca} }{ \mathrm{\Delta^{44/40}Ca}-\mathrm{\Delta^{44/40}Ca_{inf}} } + \frac{{^{88/40}K}_\mathrm{inf}}{{^{88/40}K}_\mathrm{eq}} \right)^{-1},  \label{KSr-DELTA_Ca-D11_App}
\end{multline}
\begin{multline}
    \mathrm{\Delta^{88/86}Sr} = \mathrm{\Delta^{88/86}Sr_{inf}} + \left(\mathrm{\Delta^{88/86}Sr_{eq}}-\mathrm{\Delta^{88/86}Sr_{inf}}\right) \\
    \times \left(\frac{{^{88/40}K}_\mathrm{inf}}{{^{88/40}K}_\mathrm{eq}}\right) \left( \frac{ \mathrm{\Delta^{44/40}Ca_{eq}}-\mathrm{\Delta^{44/40}Ca} }{ \mathrm{\Delta^{44/40}Ca}-\mathrm{\Delta^{44/40}Ca_{inf}} } + \frac{{^{88/40}K}_\mathrm{inf}}{{^{88/40}K}_\mathrm{eq}} \right)^{-1}.  \label{DELTA_Sr-DELTA_Ca-D11_App}
\end{multline}
\end{linenomath} 
Since the isotope fractionations are small (i.e., within a few permil), the difference between $^{88/40}K$ and $K$ is also small. For this reason, we can use $K_\mathrm{inf}/K_\mathrm{eq}$ to replace ${^{88/40}K}_\mathrm{inf}/{^{88/40}K}_\mathrm{eq}$ for the above equations (given that $K$ may vary from ${\sim}0.025$ to ${\sim}0.25$, changing its estimate by a few permil would only lead to negligible effects). Applying this replacement, and given the objectives discussed in the main text (i.e., using the measured $\mathrm{\Delta^{44/40}Ca}$ to predict $K$ and $\mathrm{\Delta^{88/86}Sr}$ under the D11 framework, which would lead to the replacement of $\mathrm{\Delta^{44/40}Ca}$, $K$, and  $\mathrm{\Delta^{88/86}Sr}$ by $\mathrm{\Delta^{44/40}Ca_{M}}$, $K_\mathrm{D11}$, and $\mathrm{\Delta^{88/86}Sr_{D11}}$), we obtain Eqs.~\ref{KSr-DELTA_Ca-D11}--\ref{DELTA_Sr-DELTA_Ca-D11} from Eqs.~\ref{KSr-DELTA_Ca-D11_App}--\ref{DELTA_Sr-DELTA_Ca-D11_App}.

\subsubsection{Ion-by-Ion model}  \label{app:ion-by-ion}
In the Ion-by-Ion model, ${^{88/40}K}_\mathrm{f}$ is constant while ${^{88/40}K}_\mathrm{b}$ is not \citep{nielsen2013general, jia2022model}. The $\mathrm{^{88}Sr}/\mathrm{^{40}Ca}$ partitioning (Eq.~\ref{Kp_88/40}) in the kinetic limit ($^{40}R_\mathrm{f}/^{40}R_\mathrm{b} \to \infty$), again, leads to Eq.~\ref{Kf_D11}; Substituting it into Eq.~\ref{Kp_88/40} leads to
\begin{linenomath} 
\begin{equation}
    {^{88/40}K} = \frac{ {^{88/40}K}_\mathrm{inf} ({^{40}R}_\mathrm{f}/{^{40}R}_\mathrm{b})}{ {^{40}R}_\mathrm{f}/{^{40}R}_\mathrm{b} + ({^{88/40}K}_\mathrm{b}-1) },  \label{Kp_88/40_ibi} 
\end{equation}
\end{linenomath} 
To test the model against the observed correlation of $K$, $\mathrm{\Delta^{44/40}Ca}$, and $\mathrm{\Delta^{88/86}Sr}$, we combine Eqs.~\ref{DELTA_44/40_final}, \ref{Kp_88/40_ibi} and \ref{DELTA_88/86_final} to eliminate ${^{40}R}_\mathrm{p}/{^{40}R}_\mathrm{b}$ and ${^{88/40}K}_\mathrm{b}$, such that we can express $\mathrm{\Delta^{88/86}Sr}$ as a function of $\mathrm{\Delta^{44/40}Ca}$ and ${^{88/40}K}$ through the following relation,
\begin{linenomath} 
\begin{multline}
    \mathrm{\Delta^{88/86}Sr} = \mathrm{\Delta^{88/86}Sr_{inf}} + \left(\mathrm{\Delta^{88/86}Sr_{eq}}-\mathrm{\Delta^{88/86}Sr_{inf}}\right) \\
    \times \left[1 + \frac{{^{88/40}K}}{{^{88/40}K}_\mathrm{inf}} \left(\frac{\mathrm{\Delta^{44/40}Ca}-\mathrm{\Delta^{44/40}Ca_{eq}}}{\mathrm{\Delta^{44/40}Ca_{eq}}-\mathrm{\Delta^{44/40}Ca_{inf}}} \right) \right].  \label{DELTA_Sr-DELTA_Ca-ibi_App}
\end{multline}
\end{linenomath} 
When deriving the above equation, we used the $\mathrm{^{88}Sr}/\mathrm{^{40}Ca}$ partition coefficient, ${^{88/40}K}$, instead of the overall Sr/Ca partition coefficient, $K$. Mathematically, these two values are not strictly identical due to the permil-level isotope fractionation effects, but their relative difference should be on the order of ${\sim}0.001$. In this circumstance, given that $K/K_\mathrm{inf}$ and $(\mathrm{\Delta^{44/40}Ca}-\mathrm{\Delta^{44/40}Ca_{eq}})/(\mathrm{\Delta^{44/40}Ca_{eq}}-\mathrm{\Delta^{44/40}Ca_{inf}})$ are both smaller than 1 in magnitude, replacing ${^{88/40}K}$ and ${^{88/40}K}_\mathrm{inf}$ with $K$ and $K_\mathrm{inf}$ would change the square-bracketed term by less than ${\sim}0.001$. Given that $\mathrm{\Delta^{88/86}Sr_{eq}}-\mathrm{\Delta^{88/86}Sr_{inf}} \approx 0.34\permil$, this replacement would alter the $\mathrm{\Delta^{88/86}Sr}$ estimate at most by ${\sim}0.0003\permil$, which is nearly two orders of magnitude smaller than the analytical uncertainty (${\sim}0.2\permil$), and thus negligible. With this substitution (i.e., $K/K_\mathrm{inf}$ for ${^{88/40}K}/{^{88/40}K}_\mathrm{inf}$), and applying the measured $\mathrm{\Delta^{44/40}Ca}$ and $K$ to predict $\mathrm{\Delta^{88/86}Sr}$ (i.e., replacing $\mathrm{\Delta^{44/40}Ca}$, $K$, and $\mathrm{\Delta^{88/86}Sr}$ by $\mathrm{\Delta^{44/40}Ca_{M}}$, $K_\mathrm{M}$, and $\mathrm{\Delta^{88/86}Sr_{IbI}}$), we obtain Eq.~\ref{DELTA_Sr-DELTA_Ca-ibi} of the main text from the above equation (Eq.~\ref{DELTA_Sr-DELTA_Ca-ibi_App}).

Here, we derive the above relation using $\mathrm{\Delta^{44/40}Ca}$ and $K$ as independent variables, instead of precisely following the detailed formulations given by \cite{nielsen2013general}, in order to maximize the usage of experimental data (since $\mathrm{\Delta^{44/40}Ca}$ and $K$ are both measured experimentally) and minimize the usage of parameters that are difficult to constrain directly through experimental and/or natural observations \citep[such as the ion detachment frequencies, although they can be inferred indirectly through fitting experimental data;][]{nielsen2013general, jia2022model}.

\subsection{A non-classical framework}  \label{app:non-classical}
Under the assumption of constant $^{44/40}\alpha_\mathrm{f}$ and $^{88/86}\alpha_\mathrm{f}$ (as well as the assumption of constant $^{44/40}\alpha_\mathrm{b}$ and $^{88/86}\alpha_\mathrm{b}$, which is not impacted by the introduction of different forward-reaction mechanisms; see Sec.~\ref{sec:new} for details), Eqs.~\ref{DELTA_44/40_final} and \ref{DELTA_88/86_final} still apply, and they together lead to
\begin{linenomath} \begin{equation}
    {^{88/40}K}_\mathrm{b} = \left( \frac{\mathrm{\Delta^{88/86}Sr}-\mathrm{\Delta^{88/86}Sr_{inf}}}{\mathrm{\Delta^{88/86}Sr_{eq}}-\mathrm{\Delta^{88/86}Sr}} \right) \left( \frac{\mathrm{\Delta^{44/40}Ca}-\mathrm{\Delta^{44/40}Ca_{inf}}}{\mathrm{\Delta^{44/40}Ca_{eq}}-\mathrm{\Delta^{44/40}Ca}} \right)^{-1}.  \label{DELTA_Sr-DELTA_Ca-Kb}
\end{equation}\end{linenomath} 

Considering the classical and non-classical forward reactions as parallel processes, we write the overall forward reaction rate of $\mathrm{^{88}Sr}$ as
\begin{linenomath} \begin{equation}
    {^{88}R}_\mathrm{f} = {^{88}R}_\mathrm{f(C)} + {^{88}R}_\mathrm{f(N)}
    = [{^{40}R}_\mathrm{f(C)}{^{88/40}K}_\mathrm{f(C)} + {^{40}R}_\mathrm{f(N)}{^{88/40}K}_\mathrm{f(N)}] (\mathrm{^{88}Sr}/\mathrm{^{40}Ca})_\mathrm{aq},
\end{equation}\end{linenomath} 
where ${^{88}R}_\mathrm{f(C)}$ and ${^{88}R}_\mathrm{f(N)}$ are the $\mathrm{^{88}Sr}$ attachment rates of classical and non-classical crystallization mechanisms, ${^{40}R}_\mathrm{f(C)}$ and ${^{40}R}_\mathrm{f(N)}$ are the $\mathrm{^{40}Ca}$ attachment rates of $\mathrm{^{40}Ca}$ associated with the two mechanisms, and ${^{88/40}K}_\mathrm{f(C)}$ and ${^{88/40}K}_\mathrm{f(N)}$ are the forward-reaction $\mathrm{^{88}Sr}/\mathrm{^{40}Ca}$ partition coefficients associated with the two mechanisms,
\begin{linenomath} \begin{equation}
    {^{88/40}K}_\mathrm{f(C)} = \frac{^{88}R_\mathrm{f(C)}/^{40}R_\mathrm{f(C)}} {(\mathrm{^{88}Sr}/\mathrm{^{40}Ca})_\mathrm{aq}},  \quad
    {^{88/40}K}_\mathrm{f(N)} = \frac{^{88}R_\mathrm{f(N)}/^{40}R_\mathrm{f(N)}} {(\mathrm{^{88}Sr}/\mathrm{^{40}Ca})_\mathrm{aq}}.  \label{Kfm_Kfp_88/40}
\end{equation}\end{linenomath} 
The forward-reaction $\mathrm{^{88}Sr}/\mathrm{^{40}Ca}$ partition coefficient is thus
\begin{linenomath} \begin{equation}
    {^{88/40}K}_\mathrm{f} = \frac{^{88}R_\mathrm{f}/^{40}R_\mathrm{f}} {(\mathrm{^{88}Sr}/\mathrm{^{40}Ca})_\mathrm{aq}}
    = {^{88/40}K}_\mathrm{f(C)} f_\mathrm{C} + {^{88/40}K}_\mathrm{f(N)} f_\mathrm{N}.  \label{88/40Kf_ZW23_prep}
\end{equation}\end{linenomath} 
where $f_\mathrm{C} = {^{40}R}_\mathrm{f(C)}/{^{40}R}_\mathrm{f}$ and $f_\mathrm{N} = {^{40}R}_\mathrm{f(N)}/{^{40}R}_\mathrm{f}$ are the fractional contributions of the classical and non-classical crystallization mechanisms ($f_\mathrm{C} + f_\mathrm{N} = 1$). Substituting Eq.~\ref{88/40Kp_ZW23_prep} into Eq.~\ref{Kp_88/40} leads to
\begin{linenomath} \begin{equation}
    {^{88/40}K} 
    = \left( {^{88/40}K}_\mathrm{f(C)} f_\mathrm{C} + {^{88/40}K}_\mathrm{f(N)} f_\mathrm{N} \right)  \frac{ ({^{40}R}_\mathrm{f}/{^{40}R}_\mathrm{b})}{ {^{40}R}_\mathrm{f}/{^{40}R}_\mathrm{b} + ({^{88/40}K}_\mathrm{b}-1) }.  \label{88/40Kp_ZW23_prep}
\end{equation}\end{linenomath} 
Applying Eq.~\ref{DELTA_44/40_final} and Eq.~\ref{DELTA_Sr-DELTA_Ca-Kb} to eliminate the reaction rates, we obtain
\begin{linenomath} \begin{equation}
    {^{88/40}K}^\prime
    = {^{88/40}K}_\mathrm{f(C)} f_\mathrm{C} + {^{88/40}K}_\mathrm{f(N)} f_\mathrm{N},  \label{Kp_KfC_KfN_App}
\end{equation}\end{linenomath} 
where ${^{88/40}K}^\prime$ is defined as
\begin{linenomath} \begin{equation}
    {^{88/40}K}^\prime = {^{88/40}K} \left( \frac{\mathrm{\Delta^{88/86}Sr_{eq}}-\mathrm{\Delta^{88/86}Sr}}{\mathrm{\Delta^{88/86}Sr_{eq}}-\mathrm{\Delta^{88/86}Sr_{inf}}} \right)^{-1} \left( \frac{\mathrm{\Delta^{44/40}Ca_{eq}}-\mathrm{\Delta^{44/40}Ca}}{ \mathrm{\Delta^{44/40}Ca_{eq}}-\mathrm{\Delta^{44/40}Ca_{inf}} } \right).  \label{Kprime_App}
\end{equation}\end{linenomath} 
In the limit of equilibrium (${^{40}R}_\mathrm{f}/{^{40}R}_\mathrm{b} = 1$), the forward reaction likely occurs through the classical crystallization mechanism (ion-by-ion attachment) only, $f_\mathrm{C} \approx 1$ and $f_\mathrm{N} \approx 0$. In the kinetic limit (${^{40}R}_\mathrm{f}/{^{40}R}_\mathrm{b} \gg 1$, which corresponds to large saturation states), the contribution of non-classical crystallization mechanism (polymer/particle attachment) is important and probably dominant; as an extreme end-member, we assume $f_\mathrm{C} \approx 0$ and $f_\mathrm{N} \approx 1$. Considering Eq.~\ref{88/40Kp_ZW23_prep} in these two limits, we obtain
\begin{linenomath} \begin{equation}
    {^{88/40}K}^\prime_\mathrm{eq} = {^{88/40}K}_\mathrm{f(C)},  \quad
    {^{88/40}K}_\mathrm{inf} = {^{88/40}K}_\mathrm{f(N)},  \label{88/40K_eq_inf_ZW23}  
\end{equation}\end{linenomath} 
where ${^{88/40}K}^\prime_\mathrm{eq} = {^{88/40}K}_\mathrm{eq} {^{88/40}K}_\mathrm{b(eq)}$ is the adjusted equilibrium $\mathrm{{^{88}Sr}/{^{40}Ca}}$ partition coefficient, and ${^{88/40}K}_\mathrm{b(eq)}$ is the backward-reaction partition coefficient at equilibrium conditions. Substituting Eq.~\ref{88/40K_eq_inf_ZW23} into Eq.~\ref{Kp_KfC_KfN_App}, we obtain the fractional contributions of the classical and non-classical crystallization mechanisms,
\begin{linenomath} \begin{equation}
    f_\mathrm{C} = \frac{ {^{88/40}K}_\mathrm{inf} - {^{88/40}K}^\prime}{ {^{88/40}K}_\mathrm{inf}-{^{88/40}K}^\prime_\mathrm{eq}}, 
 \label{fC_App} 
\end{equation}
\begin{equation}
    f_\mathrm{N} = \frac{ {^{88/40}K}^\prime - {^{88/40}K}^\prime_\mathrm{eq}}{ {^{88/40}K}_\mathrm{inf}-{^{88/40}K}^\prime_\mathrm{eq}}.  \label{fN_App}
\end{equation}\end{linenomath} 
Applying Eq.~\ref{DELTA_44/40_final}, together with $^{40}R_\mathrm{p} = {^{40}R}_\mathrm{f} - {^{40}R}_\mathrm{b}$, we express the forward reaction rate using the observed net precipitation rate and Ca isotope fractionation factor as
\begin{linenomath} \begin{equation}
    {^{40}R}_\mathrm{f} = {^{40}R}_\mathrm{p} \times \left(\frac{ \mathrm{\Delta^{44/40}Ca_{inf}}-\mathrm{\Delta^{44/40}Ca_{eq}}}{\mathrm{\Delta^{44/40}Ca}-\mathrm{\Delta^{44/40}Ca_{eq}}}\right).  \label{Rf_App}
\end{equation}\end{linenomath} 
As mentioned above, we can replace $^{88/40}K$ terms with the common Sr/Ca partition coefficient $K$ terms and consider $R_\mathrm{f}/R_\mathrm{p} \approx {^{40}R}_\mathrm{f}/{^{40}R}_\mathrm{p}$; given the purpose of this study (i.e., to quantity the contributions of the classical and non-classical crystallization pathways), we replace $\mathrm{\Delta^{44/40}Ca}$, $\mathrm{\Delta^{88/86}Sr}$ and $K$ with their measured values, $\mathrm{\Delta^{44/40}Ca_{M}}$, $\mathrm{\Delta^{88/86}Sr_{M}}$ and $K_\mathrm{M}$. With these replacements, we obtain Eqs.~\ref{fCfN}--\ref{forward_rate} of the main text from Eqs.~\ref{DELTA_Sr-DELTA_Ca-Kb}, \ref{Kprime_App}, and \ref{fC_App}--\ref{Rf_App} of this appendix.

\section{A remark on D11 ``Model 2''}  \label{app:D11M2}
As outlined in Eqs.~11 and 13 of \cite{depaolo2011surface} and detailed in our \ref{app:D11}, the D11 model was developed with $R_\mathrm{p}/R_\mathrm{b}$ (or $R_\mathrm{f}/R_\mathrm{b} = 1+R_\mathrm{p}/R_\mathrm{b}$) as the independent variable. To facilitate comparison with experimental findings concerning the ``rate dependence'', where the net precipitation rate $R_\mathrm{p}$ is treated as the independent variable, assumptions regarding $R_\mathrm{b}$ are required. In \cite{depaolo2011surface}, two types of parameterizations (``Model 1'' and ``Model 2'') were presented. Here, we primarily focus on the mutual correlations of $\mathrm{\Delta^{88/86}Sr}$, $\mathrm{\Delta^{44/40}Ca}$, and $\mathrm{K}$, for which the rate dependence is not explicitly involved, and the choice of Model 1 or Model 2 does not impact the conclusion. In this appendix, we provide some details to demonstrate this point. For calculations shown in this appendix, given that the isotope fractionation is essentially small, we replace ${^{40}R}_\mathrm{f}/{^{40}R}_\mathrm{b}$ with $R_\mathrm{f}/R_\mathrm{b}$ for Eqs.~\ref{DELTA_44/40_final}, \ref{Kp_88/40_D11}, and \ref{DELTA88/86_D11}. With this practice, our equations for Ca isotope fractionation and Sr/Ca partitioning (Eqs.~\ref{DELTA_44/40_final} and \ref{Kp_88/40_D11}) can be shown to be equivalent to those provided by \cite{depaolo2011surface} in his Eqs. 11 and 13.

In the general case of the D11 model framework, we calculated the variations of $\mathrm{\Delta^{44/40}Ca}$, $K$, and $\mathrm{\Delta^{88/86}Sr}$ with $R_\mathrm{p}/R_\mathrm{b}$ (Figs.~\ref{fig:D11M2}a--c), following Eqs.~\ref{DELTA_44/40_final}, \ref{Kp_88/40_D11}, and \ref{DELTA88/86_D11} (note that $R_\mathrm{p}/R_\mathrm{b} = R_\mathrm{f}/R_\mathrm{b}-1$). For these equations, no assumption was made for $R_\mathrm{b}$, and $R_\mathrm{p}/R_\mathrm{b}$ cannot be converted to $R_\mathrm{p}$. In this case, comparison between model predictions and observations cannot be made for $\mathrm{\Delta^{44/40}Ca}$--$R_\mathrm{p}$, $K$--$R_\mathrm{p}$, and $\mathrm{\Delta^{88/86}Sr}$--$R_\mathrm{p}$ relations. However, we can demonstrate using these results that the D11 model framework is inadequate to explain the $\mathrm{\Delta^{88/86}Sr}$--$\mathrm{\Delta^{44/40}Ca}$ correlation (Fig.~\ref{fig:D11M2}d), as discussed in Sec.~\ref{sec:previous}. 

For D11 ``Model 1'' (Figs.~\ref{fig:D11M2}e--i), $R_\mathrm{b}$ is held as a constant,
\begin{linenomath} \begin{equation}
    R_\mathrm{b} = 6\times 10^{-7}\ \mathrm{mol/m^2/s},  \label{M1Rb}
\end{equation}\end{linenomath} 
as chosen by \cite{depaolo2011surface}, following the calcite dissolution experiments of \cite{chou1989comparative}, and shown in Fig.~\ref{fig:D11M2}e. With this parameterization, the results fit the $\mathrm{\Delta^{44/40}Ca}$--$R_\mathrm{p}$ data relatively well; however, they do not fit the $K$--$R_\mathrm{p}$ and $\mathrm{\Delta^{88/86}Sr}$--$R_\mathrm{p}$ data (Figs.~\ref{fig:D11M2}g and h). We note that \cite{depaolo2011surface} assumed an equilibrium Sr/Ca partition coefficient of $K_\mathrm{eq} = 0.07$, which would yield a better fitting of $K$--$R_\mathrm{p}$ (as shown in his Fig.~10a); however, as the recent study of marine sediment and pore fluids \citep{zhang2020equilibrium} suggested an equilibrium Sr/Ca partition coefficient of $K_\mathrm{eq} = 0.025$, we perform our calculations here with $K_\mathrm{eq} = 0.025$ instead of $K_\mathrm{eq} = 0.07$ and obtained the results shown in our Fig.~\ref{fig:D11M2}. Consequently, the predictions of D11 Model 1 cannot account for the experimentally observed $\mathrm{\Delta^{44/40}Ca}$--$\mathrm{\Delta^{88/86}Sr}$ correlation (Fig.~\ref{fig:D11M2}i, which is exactly same as Fig.~\ref{fig:D11M2}d).

To better account for the experimental $\mathrm{\Delta^{44/40}Ca}$--$R_\mathrm{p}$ data, \cite{depaolo2011surface} introduced a ``Model 2'' for the parameterization of $R_\mathrm{b}$, which can be expressed as
\begin{linenomath} \begin{equation}
    R_\mathrm{b} = \min\left(\Lambda {R_\mathrm{p}}^{1/2}, 6\times 10^{-7}\ \mathrm{mol/m^2/s}\right).  \label{M2Rb}
\end{equation}\end{linenomath} 
This expression implies that the backward reaction rate deceases as the system approaches the equilibrium state, which is consistent with predictions of the Ion-by-Ion model \citep{nielsen2012self}. We first show the results with 
\begin{linenomath} \begin{equation}
    \Lambda = 8\times10^{-4}\ \mathrm{(mol/m^2/s)^{1/2}},  \label{M2C1Rb} 
\end{equation}\end{linenomath} 
which leads to a $R_\mathrm{b}$--$R_\mathrm{p}$ relation (Figs.~\ref{fig:D11M2}j) that is same as the one used by \cite{depaolo2011surface} to fit the $\mathrm{\Delta^{44/40}Ca}$--$R_\mathrm{p}$ data. In this specific example (referred to here as Model 2 Case 1, see Figs.~\ref{fig:D11M2}j--n), the results fit the $\mathrm{\Delta^{44/40}Ca}$--$R_\mathrm{p}$ data particularly well (Fig.~\ref{fig:D11M2}k), even better than D11 ``Model 1'' (Fig.~\ref{fig:D11M2}f). However, the predicted curves still do not explain the $K$--$R_\mathrm{p}$ and $\mathrm{\Delta^{88/86}Sr}$--$R_\mathrm{p}$ data (Figs.~\ref{fig:D11M2}l and m), and the predictions still cannot account for the $\mathrm{\Delta^{44/40}Ca}$--$\mathrm{\Delta^{88/86}Sr}$ correlation (Fig.~\ref{fig:D11M2}n, which is same as Figs.~\ref{fig:D11M2}d and i).

We also adjust the $R_\mathrm{b}$--$R_\mathrm{p}$ relation with (Figs.~\ref{fig:D11M2}o)
\begin{linenomath} \begin{equation}
    \Lambda = 1\times10^{-4}\ \mathrm{(mol/m^2/s)^{1/2}},  \label{M2C2Rb} 
\end{equation}\end{linenomath} 
which is the same to the one used by \cite{zhang2020equilibrium} to explain the $K$--$R_\mathrm{p}$ data of the \cite{tang2008I} experiments \citep[their samples were also meausred by][and reported together with Sr isotope data]{bohm2012strontium}. In this specific case (referred as ``Model 2 Case 2'', see Figs.~\ref{fig:D11M2}o--s), the results fit the $K$--$R_\mathrm{p}$ and $\mathrm{\Delta^{88/86}Sr}$--$R_\mathrm{p}$ data well (Figs.~\ref{fig:D11M2}q and r), but they do not fit the $\mathrm{\Delta^{44/40}Ca}$--$R_\mathrm{p}$ data (Figs.~\ref{fig:D11M2}p), and thus still cannot explain the $\mathrm{\Delta^{44/40}Ca}$--$\mathrm{\Delta^{88/86}Sr}$ correlation (Fig.~\ref{fig:D11M2}s, which is same as Figs.~\ref{fig:D11M2}d, i, and n). 

All these calculations demonstrate that, with any specific parameterization of $R_\mathrm{b}$ (including D11 ``Model 1'', D11 ``Model 2 Case 1'', or D11 ``Model 2 Case 2''), the D11 predictions cannot simultaneously explain the $\mathrm{\Delta^{44/40}Ca}$--$R_\mathrm{p}$ and $\mathrm{\Delta^{88/86}Sr}$--$R_\mathrm{p}$ observations, nor the $\mathrm{\Delta^{88/86}Sr}$--$\mathrm{\Delta^{44/40}Ca}$ correlation, as shown in Figs.~\ref{fig:D11M2}d, i, n, and s, which are exactly the same because the predictions using the D11 framework on $\mathrm{\Delta^{88/86}Sr}$--$\mathrm{\Delta^{44/40}Ca}$ correlation are not affected by $R_\mathrm{b}$ parameterization as discussed in the main text.

\clearpage
\begin{figure}[ht!]
\centering
    \includegraphics[width=1.0\textwidth]{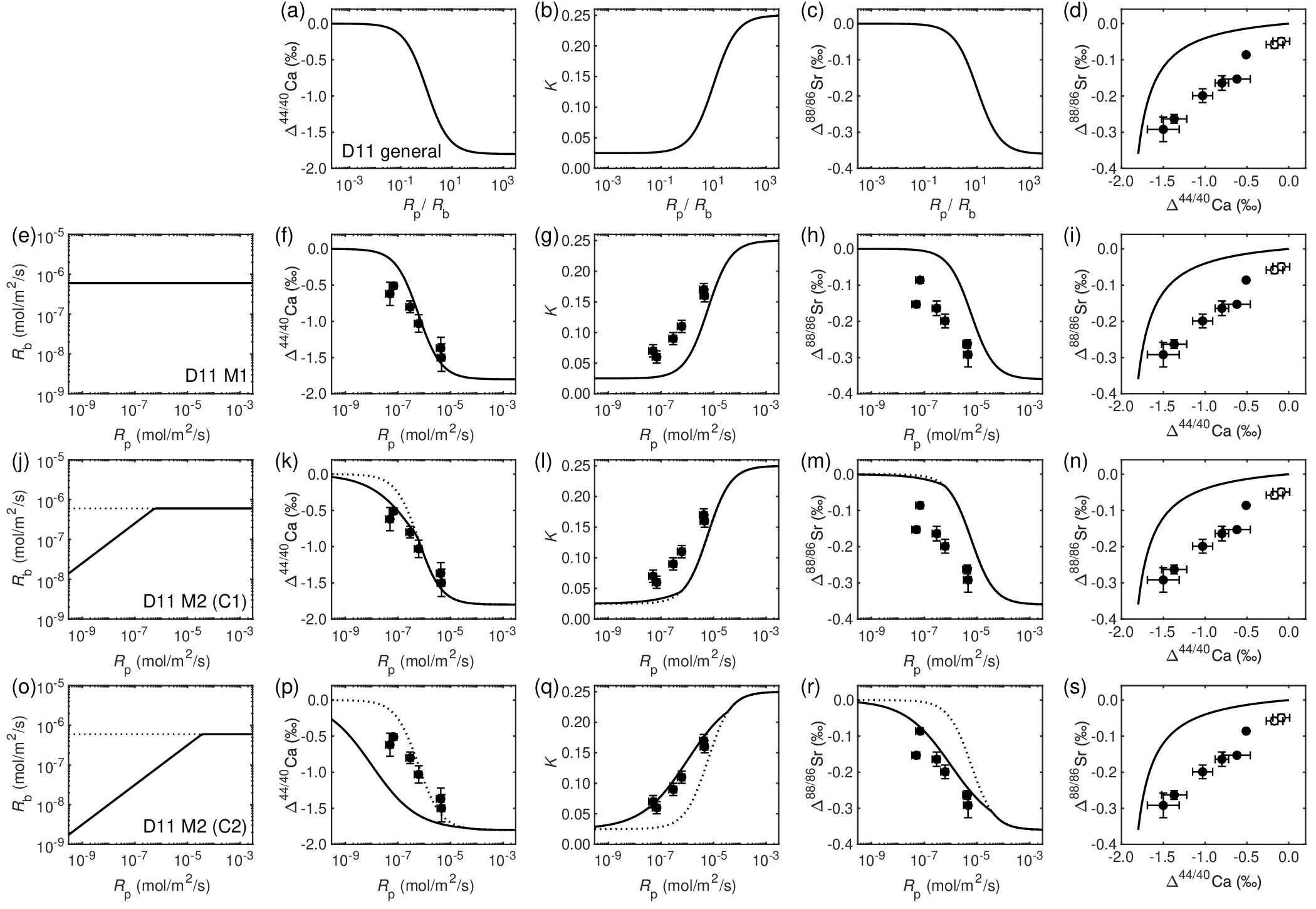}
    \caption{The comparison of predictions under the D11 framework with experimental observations. Synthetic and ocean crust calcites are illustrated by solid circles and open circles, respectively \citep[][]{bohm2012strontium}. (a--d) The general D11 model using $R_\mathrm{p}/R_\mathrm{b}$ as independent variable. Comparison is only made for the $\mathrm{\Delta^{88/86}Sr}$--$\mathrm{\Delta^{44/40}Ca}$ correlation (d) because $R_\mathrm{p}/R_\mathrm{b}$ is not directly observable; (e--i) D11 ``Model 1'' predictions using $R_\mathrm{b}$ given by Eq.~\ref{M1Rb} shown in panel (e); (j--n) A specific case (Case 1) of D11 Model 2, in which $R_\mathrm{b}$ varies with $R_\mathrm{p}$ as shown in Eqs.~\ref{M2Rb} and \ref{M2C1Rb}, as assumed in \cite{depaolo2011surface} and shown in panel (j). (o--s) A specific case (Case 2) of D11 ``Model 2'', in which $R_\mathrm{b}$ varies with $R_\mathrm{p}$ following Eqs.~\ref{M2Rb} and \ref{M2C2Rb}, as assumed in \cite{zhang2020equilibrium} and shown in panel (o); In (j--s), the results of D11 ``Model 1'' are shown in dotted curves for comparison. For all different choices for $R_\mathrm{b}$ (Model 1, Model 2 Case 1, and Model 2 Case 2), under the D11 framework, the predicted $\mathrm{\Delta^{88/86}Sr}$--$\mathrm{\Delta^{44/40}Ca}$ correlations are exactly identical as shown in panels (d, i, n, s).} \label{fig:D11M2}
\end{figure}

\section{The effects of adjusting kinetic-limit parameters}  \label{app:parameters}
The root-square-mean error for $\mathrm{\Delta^{88/86}Sr}$ predicted by the D11 model (in comparison with the experimental measurements) is calculated as
\begin{equation}
    \mathrm{RSME(D11)} = \sqrt{ \frac{1}{N} \sum_{i=1}^N \Big[\mathrm{\Delta^{88/86}Sr_{D11}(i)}-\mathrm{\Delta^{88/86}Sr_{M}(i)} \Big]^2 },
\end{equation}
where $\mathrm{\Delta^{88/86}Sr_{M}(i)}$ is the measured value of Sr fractionation factor of the $i$-th sample of \cite{bohm2012strontium}, $\mathrm{\Delta^{88/86}Sr_{D11}(i)}$ is the Sr fractionation factor predicted using the D11 model framework (using Eq.~\ref{DELTA_Sr-DELTA_Ca-D11} of the main text), and $N$ is the number of measurements. Similarly, the root-square-mean error for $\mathrm{\Delta^{88/86}Sr}$ predicted by the Ion-by-Ion model (in comparison with the experimental measurements) is calculated as
\begin{equation}
    \mathrm{RSME(ion{\text -}by{\text -}ion)} = \sqrt{ \frac{1}{N} \sum_{i=1}^N \Big[\mathrm{\Delta^{88/86}Sr_{IbI}(i)}-\mathrm{\Delta^{88/86}Sr_{M}(i)} \Big]^2 },
\end{equation}
where $\mathrm{\Delta^{88/86}Sr_{IbI}(i)}$ is the Sr fractionation factor predicted using the Ion-by-Ion model framework (using Eq.~\ref{DELTA_Sr-DELTA_Ca-ibi} of the main text).

Apart from $\mathrm{\Delta^{44/40}Ca_{M}}$ and $K_\mathrm{M}$, the calculations of $\mathrm{\Delta^{88/86}Sr_{D11}}$ and $\mathrm{\Delta^{88/86}Sr_{IbI}}$ also involve six parameters: three for the equilibrium limit, $\mathrm{\Delta^{44/40}Ca}_\mathrm{eq}$, $\mathrm{\Delta^{88/86}Sr}_\mathrm{eq}$, and $K_\mathrm{eq}$, and three for the kinetic limit, $\mathrm{\Delta^{44/40}Ca}_\mathrm{inf}$, $\mathrm{\Delta^{88/86}Sr_{inf}}$, and $K_\mathrm{inf}$ (Eqs.~\ref{DELTA_Sr-DELTA_Ca-D11} and \ref{DELTA_Sr-DELTA_Ca-ibi} of the main text). In this appendix, we explore the effects of choosing different values of these parameters (within the reasonable range) on the the root-square-mean errors of the predictions of the two models. We mainly discuss parameters for the kinetic limit, because the equilibrium limit has been well constrained by natural and experimental observations. The $K_\mathrm{inf}$ value is well constrained by experimental results \citep{gabitov2006partitioning} to be in the range of 0.25--0.35, and we only consider three values, 0.25, 0.30, and 0.35. The values of $\mathrm{\Delta^{44/40}Ca}_\mathrm{inf}$ and $\mathrm{\Delta^{88/86}Sr_{inf}}$ are less well constrained and, thus, are the focus of the following discussion. 

The values of $\mathrm{\Delta^{44/40}Ca}$ and $\mathrm{\Delta^{88/86}Sr}$ both decrease with increasing precipitation rate \citep{tang2008II, bohm2012strontium}. Hence, $\mathrm{\Delta^{44/40}Ca}_\mathrm{inf}$ and $\mathrm{\Delta^{88/86}Sr_{inf}}$ should be at least larger-in-magnitude than largest fractionation obtained in the experiments \citep[i.e., smaller than approximately $-1.5\permil$ for $\mathrm{\Delta^{44/40}Ca}$ and $-0.29\permil$ for $\mathrm{\Delta^{88/86}Sr}$;][]{tang2008II, bohm2012strontium}. We change these parameters within the reasonable ranges described above, and calculate the root-mean-square errors of the predictions of the D11 model and Ion-by-Ion model (Fig.~\ref{fig:PARAMdSr00}). Within the explored parameter space, the root-mean-square error for $\mathrm{\Delta^{88/86}Sr}$ is always larger than ${\sim}0.07\permil$ for the D11 model and ${\sim}0.06\permil$ for the Ion-by-Ion model, which are considerably greater than the root-mean-square error of our model prediction, ${\sim}0.02\permil$. Moreover, the minimum root-square-mean errors, for both the D11 model and Ion-by-Ion model, are approached only at high rather high $\mathrm{\Delta^{88/86}Sr_{inf}}/\mathrm{\Delta^{44/40}Ca}_\mathrm{inf}$ ratios (i.e., in the lower-right corners of the figures); theoretical considerations of the isotope fractionation process suggests these values are out of the reasonable range of the parameters.

Now, we discuss the theoretical constraint on the relation of $\mathrm{\Delta^{44/40}Ca}_\mathrm{inf}$ and $\mathrm{\Delta^{88/86}Sr_{inf}}$. Assuming the forward-reaction isotope fractionation is caused by the dehydration of ions \citep[e.g.,][]{depaolo2011surface, hofmann2012ion}, the forward-reaction Ca and Sr isotope fractionation factors can be expressed by
\begin{equation}
    {^{44/40}\alpha_\mathrm{f}} = (44/40)^{-\gamma} \approx 1 - 0.1\gamma, \quad
    {^{88/86}\alpha_\mathrm{f}} = (88/86)^{-\gamma} \approx 1 - 0.02\gamma,  \label{alpha-inf}
\end{equation}
where $\gamma$ is a constant coefficient \citep[e.g.,][]{hofmann2012ion}, and ${^{44/40}\alpha_\mathrm{f}}$ and ${^{88/86}\alpha_\mathrm{f}}$ are the forward-reaction Ca and Sr isotope fractionation factors (also defined in the Methods section of the paper). These two factors can be related to $\mathrm{\Delta^{44/40}Ca}_\mathrm{inf}$ and $\mathrm{\Delta^{88/86}Sr_{inf}}$ as
\begin{equation}
    \mathrm{\Delta^{44/40}Ca_{inf}} = 1000\permil \times ({^{44/40}\alpha_\mathrm{f}}-1), \quad
    \mathrm{\Delta^{88/86}Sr_{inf}} = 1000\permil \times ({^{88/86}\alpha_\mathrm{f}}-1), \label{DELTA-alpha-inf}
\end{equation}
which can be obtained from Eqs.~18 and 26 of the main text. Eqs.~\ref{alpha-inf} and \ref{DELTA-alpha-inf} suggest that 
\begin{equation}
    \mathrm{\Delta^{88/86}Sr_{inf}}/\mathrm{\Delta^{44/40}Ca_{inf}} \approx 0.2. \label{DELTA-ratio}
\end{equation}
This additional theoretical constraint (Eq.~\ref{DELTA-ratio}) is labeled in the figures by the dashed lines (Fig.~\ref{fig:PARAMdSr00}). When this theoretical constraint is satisfied (i.e., along the dashed lines in the figures), the root-mean-square error for $\mathrm{\Delta^{88/86}Sr}$ remains greater than ${\sim}0.12\permil$ for the D11 model and ${\sim}0.09\permil$ for the Ion-by-Ion model, suggesting the undesirable fits for both models. 

As discussed in Sec.~\ref{sec:previous}, although we prefer an equilibrium Sr isotope fractionation factor of $\mathrm{\Delta^{88/86}Sr_{eq}} = 0\permil$ (see \ref{app:limit_parameters}), we comprehend that the direct observation provided by \cite{bohm2012strontium} can only confirm that $\mathrm{\Delta^{88/86}Sr_{eq}}$ must be smaller than $-0.05\permil$ in magnitude. Here, we also perform calculations with $\mathrm{\Delta^{88/86}Sr_{eq}} = -0.05\permil$ (Fig.~\ref{fig:PARAMdSr05}). With this adjustment, when the theoretical constraint of Eq.~\ref{DELTA-ratio} is satisfied (i.e., along the dashed lines in the figures), the root-mean-square error for $\mathrm{\Delta^{88/86}Sr}$ remains greater than ${\sim}0.09\permil$ for the D11 model and ${\sim}0.05\permil$ for the Ion-by-Ion model, still considerably larger than the root-mean-square error of our model prediction and greater than the analytical uncertainty.

\begin{figure}[ht!]
\centering
    \includegraphics[width=1.0\textwidth]{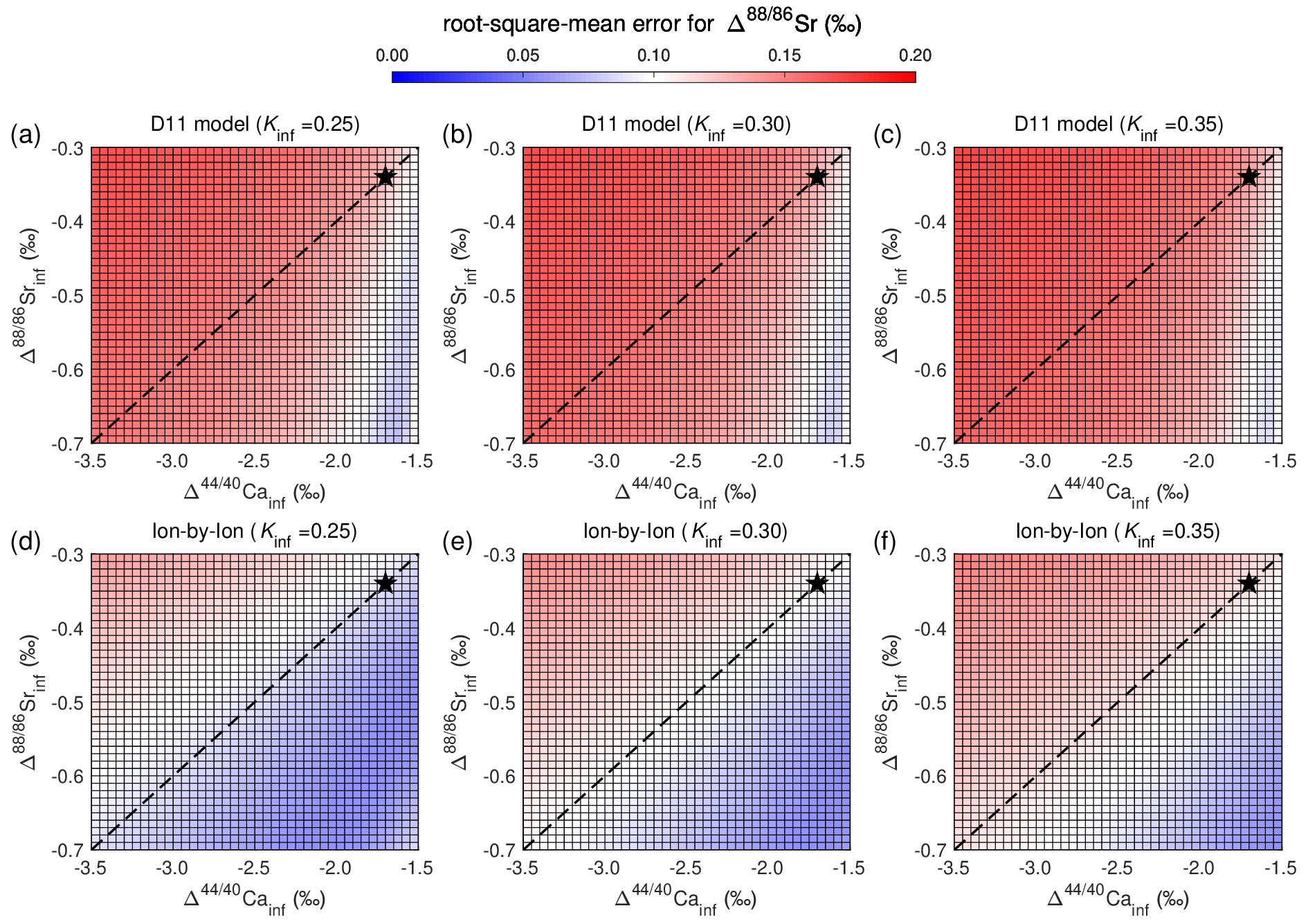}
    \caption{The root-square-mean error for $\mathrm{\Delta^{88/86}Sr}$ predicted by the D11 model (a--c) and Ion-by-Ion model (d--f) with different choices of kinetic-limit parameters. The dashed lines indicate the $\mathrm{\Delta^{88/86}Sr_{inf}}/\mathrm{\Delta^{44/40}Ca_{inf}}$ ratio suggested by theoretical study \citep{hofmann2012ion}. The stars indicate $\mathrm{\Delta^{88/86}Sr_{inf}}$ and $\mathrm{\Delta^{44/40}Ca_{inf}}$ assumed in the main text. The calculations are performed with the preferred values of equilibrium parameters described in \ref{app:limit_parameters}}  \label{fig:PARAMdSr00}
\end{figure}

\begin{figure}[ht!]
\centering
    \includegraphics[width=1.0\textwidth]{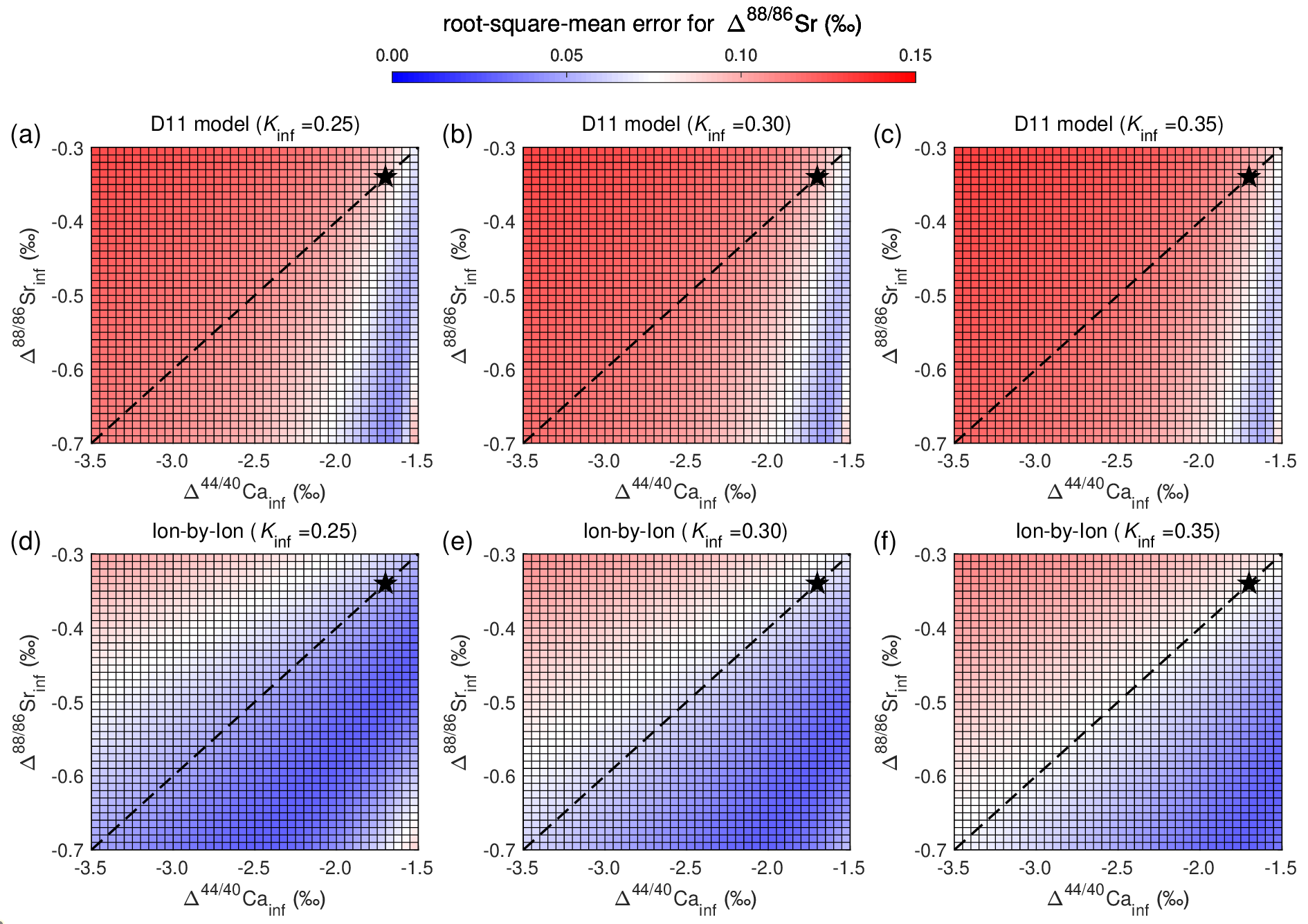}
    \caption{Same as Fig.~\ref{fig:PARAMdSr00} but with calculations performed assuming $\mathrm{\Delta^{88/86}Sr_\mathrm{eq}} = -0.05\permil$.}  \label{fig:PARAMdSr05}
\end{figure}

\clearpage
\bibliographystyle{elsarticle-harv} 
\bibliography{cas-refs}

\begin{thebibliography}{59}
\expandafter\ifx\csname natexlab\endcsname\relax\def\natexlab#1{#1}\fi
\providecommand{\url}[1]{\texttt{#1}}
\providecommand{\href}[2]{#2}
\providecommand{\path}[1]{#1}
\providecommand{\DOIprefix}{doi:}
\providecommand{\ArXivprefix}{arXiv:}
\providecommand{\URLprefix}{URL: }
\providecommand{\Pubmedprefix}{pmid:}
\providecommand{\doi}[1]{\href{http://dx.doi.org/#1}{\path{#1}}}
\providecommand{\Pubmed}[1]{\href{pmid:#1}{\path{#1}}}
\providecommand{\bibinfo}[2]{#2}
\ifx\xfnm\relax \def\xfnm[#1]{\unskip,\space#1}\fi
\bibitem[{AlKhatib and Eisenhauer(2017)}]{alkhatib2017calcium}
\bibinfo{author}{AlKhatib, M.}, \bibinfo{author}{Eisenhauer, A.}, \bibinfo{year}{2017}.
\newblock \bibinfo{title}{Calcium and strontium isotope fractionation in aqueous solutions as a function of temperature and reaction rate; {I}. calcite}.
\newblock \bibinfo{journal}{Geochimica et Cosmochimica Acta} \bibinfo{volume}{209}, \bibinfo{pages}{296--319}.
\bibitem[{Andersson et~al.(2016)Andersson, Dobbersch{\"u}tz, Sand, Tobler, De~Yoreo and Stipp}]{andersson2016microkinetic}
\bibinfo{author}{Andersson, M.}, \bibinfo{author}{Dobbersch{\"u}tz, S.}, \bibinfo{author}{Sand, K.K.}, \bibinfo{author}{Tobler, D.}, \bibinfo{author}{De~Yoreo, J.J.}, \bibinfo{author}{Stipp, S.}, \bibinfo{year}{2016}.
\newblock \bibinfo{title}{A microkinetic model of calcite step growth}.
\newblock \bibinfo{journal}{Angewandte Chemie} \bibinfo{volume}{128}, \bibinfo{pages}{11252--11256}.
\bibitem[{B{\"o}hm et~al.(2012)B{\"o}hm, Eisenhauer, Tang, Dietzel, Krabbenh{\"o}ft, Kisak{\"u}rek and Horn}]{bohm2012strontium}
\bibinfo{author}{B{\"o}hm, F.}, \bibinfo{author}{Eisenhauer, A.}, \bibinfo{author}{Tang, J.}, \bibinfo{author}{Dietzel, M.}, \bibinfo{author}{Krabbenh{\"o}ft, A.}, \bibinfo{author}{Kisak{\"u}rek, B.}, \bibinfo{author}{Horn, C.}, \bibinfo{year}{2012}.
\newblock \bibinfo{title}{Strontium isotope fractionation of planktic foraminifera and inorganic calcite}.
\newblock \bibinfo{journal}{Geochimica et Cosmochimica Acta} \bibinfo{volume}{93}, \bibinfo{pages}{300--314}.
\bibitem[{Burton et~al.(1951)Burton, Cabrera and Frank}]{burton1951growth}
\bibinfo{author}{Burton, W.K.}, \bibinfo{author}{Cabrera, N.}, \bibinfo{author}{Frank, F.}, \bibinfo{year}{1951}.
\newblock \bibinfo{title}{The growth of crystals and the equilibrium structure of their surfaces}.
\newblock \bibinfo{journal}{Philosophical Transactions of the Royal Society of London. Series A, Mathematical and Physical Sciences} \bibinfo{volume}{243}, \bibinfo{pages}{299--358}.
\bibitem[{Carpenter and Lohmann(1992)}]{carpenter1992srmg}
\bibinfo{author}{Carpenter, S.J.}, \bibinfo{author}{Lohmann, K.C.}, \bibinfo{year}{1992}.
\newblock \bibinfo{title}{{Sr/Mg} ratios of modern marine calcite: Empirical indicators of ocean chemistry and precipitation rate}.
\newblock \bibinfo{journal}{Geochimica et Cosmochimica Acta} \bibinfo{volume}{56}, \bibinfo{pages}{1837--1849}.
\bibitem[{Chou et~al.(1989)Chou, Garrels and Wollast}]{chou1989comparative}
\bibinfo{author}{Chou, L.}, \bibinfo{author}{Garrels, R.M.}, \bibinfo{author}{Wollast, R.}, \bibinfo{year}{1989}.
\newblock \bibinfo{title}{Comparative study of the kinetics and mechanisms of dissolution of carbonate minerals}.
\newblock \bibinfo{journal}{Chemical geology} \bibinfo{volume}{78}, \bibinfo{pages}{269--282}.
\bibitem[{De~La~Rocha and DePaolo(2000)}]{de2000isotopic}
\bibinfo{author}{De~La~Rocha, C.L.}, \bibinfo{author}{DePaolo, D.J.}, \bibinfo{year}{2000}.
\newblock \bibinfo{title}{Isotopic evidence for variations in the marine calcium cycle over the cenozoic}.
\newblock \bibinfo{journal}{Science} \bibinfo{volume}{289}, \bibinfo{pages}{1176--1178}.
\bibitem[{De~Yoreo et~al.(2015)De~Yoreo, Gilbert, Sommerdijk, Penn, Whitelam, Joester, Zhang, Rimer, Navrotsky, Banfield et~al.}]{de2015crystallization}
\bibinfo{author}{De~Yoreo, J.J.}, \bibinfo{author}{Gilbert, P.U.}, \bibinfo{author}{Sommerdijk, N.A.}, \bibinfo{author}{Penn, R.L.}, \bibinfo{author}{Whitelam, S.}, \bibinfo{author}{Joester, D.}, \bibinfo{author}{Zhang, H.}, \bibinfo{author}{Rimer, J.D.}, \bibinfo{author}{Navrotsky, A.}, \bibinfo{author}{Banfield, J.F.}, et~al., \bibinfo{year}{2015}.
\newblock \bibinfo{title}{Crystallization by particle attachment in synthetic, biogenic, and geologic environments}.
\newblock \bibinfo{journal}{Science} \bibinfo{volume}{349}, \bibinfo{pages}{aaa6760}.
\bibitem[{DePaolo(2004)}]{depaolo2004calcium}
\bibinfo{author}{DePaolo, D.J.}, \bibinfo{year}{2004}.
\newblock \bibinfo{title}{Calcium isotopic variations produced by biological, kinetic, radiogenic and nucleosynthetic processes}.
\newblock \bibinfo{journal}{Reviews in mineralogy and geochemistry} \bibinfo{volume}{55}, \bibinfo{pages}{255--288}.
\bibitem[{DePaolo(2011)}]{depaolo2011surface}
\bibinfo{author}{DePaolo, D.J.}, \bibinfo{year}{2011}.
\newblock \bibinfo{title}{Surface kinetic model for isotopic and trace element fractionation during precipitation of calcite from aqueous solutions}.
\newblock \bibinfo{journal}{Geochimica et cosmochimica acta} \bibinfo{volume}{75}, \bibinfo{pages}{1039--1056}.
\bibitem[{Dove and Hochella(1993)}]{dove1993calcite}
\bibinfo{author}{Dove, P.M.}, \bibinfo{author}{Hochella, M.F.}, \bibinfo{year}{1993}.
\newblock \bibinfo{title}{Calcite precipitation mechanisms and inhibition by orthophosphate: In situ observations by scanning force microscopy}.
\newblock \bibinfo{journal}{Geochimica et cosmochimica acta} \bibinfo{volume}{57}, \bibinfo{pages}{705--714}.
\bibitem[{van~der Eerden(1993)}]{van1993crystal}
\bibinfo{author}{van~der Eerden, J.}, \bibinfo{year}{1993}.
\newblock \bibinfo{title}{Crystal growth mechanisms}, in: \bibinfo{booktitle}{Fundamentals}. \bibinfo{publisher}{Elsevier}, pp. \bibinfo{pages}{307--475}.
\bibitem[{Fantle and DePaolo(2005)}]{fantle2005variations}
\bibinfo{author}{Fantle, M.S.}, \bibinfo{author}{DePaolo, D.J.}, \bibinfo{year}{2005}.
\newblock \bibinfo{title}{Variations in the marine {Ca} cycle over the past 20 million years}.
\newblock \bibinfo{journal}{Earth and Planetary Science Letters} \bibinfo{volume}{237}, \bibinfo{pages}{102--117}.
\bibitem[{Fantle and DePaolo(2007)}]{fantle2007isotopes}
\bibinfo{author}{Fantle, M.S.}, \bibinfo{author}{DePaolo, D.J.}, \bibinfo{year}{2007}.
\newblock \bibinfo{title}{Ca isotopes in carbonate sediment and pore fluid from {ODP} site 807a: The {Ca$^{2+}$} (aq)--calcite equilibrium fractionation factor and calcite recrystallization rates in pleistocene sediments}.
\newblock \bibinfo{journal}{Geochimica et Cosmochimica Acta} \bibinfo{volume}{71}, \bibinfo{pages}{2524--2546}.
\bibitem[{Fietzke and Eisenhauer(2006)}]{fietzke2006determination}
\bibinfo{author}{Fietzke, J.}, \bibinfo{author}{Eisenhauer, A.}, \bibinfo{year}{2006}.
\newblock \bibinfo{title}{Determination of temperature-dependent stable strontium isotope ({$^{88}$Sr/$^{86}$Sr}) fractionation via bracketing standard {MC-ICP-MS}}.
\newblock \bibinfo{journal}{Geochemistry, Geophysics, Geosystems} \bibinfo{volume}{7}.
\bibitem[{F{\"u}ger et~al.(2022)F{\"u}ger, Kuessner, Rollion-Bard, Leis, Magna, Dietzel and Mavromatis}]{fuger2022effect}
\bibinfo{author}{F{\"u}ger, A.}, \bibinfo{author}{Kuessner, M.}, \bibinfo{author}{Rollion-Bard, C.}, \bibinfo{author}{Leis, A.}, \bibinfo{author}{Magna, T.}, \bibinfo{author}{Dietzel, M.}, \bibinfo{author}{Mavromatis, V.}, \bibinfo{year}{2022}.
\newblock \bibinfo{title}{Effect of growth rate and {pH} on {Li} isotope fractionation during its incorporation in calcite}.
\newblock \bibinfo{journal}{Geochimica et Cosmochimica Acta} .
\bibitem[{Gabitov et~al.(2014)Gabitov, Sadekov and Leinweber}]{gabitov2014crystal}
\bibinfo{author}{Gabitov, R.}, \bibinfo{author}{Sadekov, A.}, \bibinfo{author}{Leinweber, A.}, \bibinfo{year}{2014}.
\newblock \bibinfo{title}{Crystal growth rate effect on {Mg/Ca} and {Sr/Ca} partitioning between calcite and fluid: An in situ approach}.
\newblock \bibinfo{journal}{Chemical Geology} \bibinfo{volume}{367}, \bibinfo{pages}{70--82}.
\bibitem[{Gabitov and Watson(2006)}]{gabitov2006partitioning}
\bibinfo{author}{Gabitov, R.I.}, \bibinfo{author}{Watson, E.B.}, \bibinfo{year}{2006}.
\newblock \bibinfo{title}{Partitioning of strontium between calcite and fluid}.
\newblock \bibinfo{journal}{Geochemistry, Geophysics, Geosystems} \bibinfo{volume}{7}.
\bibitem[{Gratz et~al.(1993)Gratz, Hillner and Hansma}]{gratz1993step}
\bibinfo{author}{Gratz, A.}, \bibinfo{author}{Hillner, P.}, \bibinfo{author}{Hansma, P.}, \bibinfo{year}{1993}.
\newblock \bibinfo{title}{Step dynamics and spiral growth on calcite}.
\newblock \bibinfo{journal}{Geochimica et Cosmochimica Acta} \bibinfo{volume}{57}, \bibinfo{pages}{491--495}.
\bibitem[{Gussone et~al.(2005)Gussone, B{\"o}hm, Eisenhauer, Dietzel, Heuser, Teichert, Reitner, W{\"o}rheide and Dullo}]{gussone2005calcium}
\bibinfo{author}{Gussone, N.}, \bibinfo{author}{B{\"o}hm, F.}, \bibinfo{author}{Eisenhauer, A.}, \bibinfo{author}{Dietzel, M.}, \bibinfo{author}{Heuser, A.}, \bibinfo{author}{Teichert, B.M.}, \bibinfo{author}{Reitner, J.}, \bibinfo{author}{W{\"o}rheide, G.}, \bibinfo{author}{Dullo, W.C.}, \bibinfo{year}{2005}.
\newblock \bibinfo{title}{Calcium isotope fractionation in calcite and aragonite}.
\newblock \bibinfo{journal}{Geochimica et Cosmochimica Acta} \bibinfo{volume}{69}, \bibinfo{pages}{4485--4494}.
\bibitem[{Gussone and Dietzel(2016)}]{gussone2016calcium}
\bibinfo{author}{Gussone, N.}, \bibinfo{author}{Dietzel, M.}, \bibinfo{year}{2016}.
\newblock \bibinfo{title}{Calcium isotope fractionation during mineral precipitation from aqueous solution}, in: \bibinfo{booktitle}{Calcium Stable Isotope Geochemistry}. \bibinfo{publisher}{Springer}, pp. \bibinfo{pages}{75--110}.
\bibitem[{Gussone et~al.(2009)Gussone, H{\"o}nisch, Heuser, Eisenhauer, Spindler and Hemleben}]{gussone2009critical}
\bibinfo{author}{Gussone, N.}, \bibinfo{author}{H{\"o}nisch, B.}, \bibinfo{author}{Heuser, A.}, \bibinfo{author}{Eisenhauer, A.}, \bibinfo{author}{Spindler, M.}, \bibinfo{author}{Hemleben, C.}, \bibinfo{year}{2009}.
\newblock \bibinfo{title}{A critical evaluation of calcium isotope ratios in tests of planktonic foraminifers}.
\newblock \bibinfo{journal}{Geochimica et Cosmochimica Acta} \bibinfo{volume}{73}, \bibinfo{pages}{7241--7255}.
\bibitem[{Harrison et~al.(2023)Harrison, Heuser, Liebetrau, Eisenhauer, Schott and Mavromatis}]{harrison2023equilibrium}
\bibinfo{author}{Harrison, A.L.}, \bibinfo{author}{Heuser, A.}, \bibinfo{author}{Liebetrau, V.}, \bibinfo{author}{Eisenhauer, A.}, \bibinfo{author}{Schott, J.}, \bibinfo{author}{Mavromatis, V.}, \bibinfo{year}{2023}.
\newblock \bibinfo{title}{Equilibrium ca isotope fractionation and the rates of isotope exchange in the calcite-fluid and aragonite-fluid systems at {25$^\circ$ C}}.
\newblock \bibinfo{journal}{Earth and planetary science letters} \bibinfo{volume}{603}, \bibinfo{pages}{117985}.
\bibitem[{Hofmann et~al.(2012)Hofmann, Bourg and DePaolo}]{hofmann2012ion}
\bibinfo{author}{Hofmann, A.E.}, \bibinfo{author}{Bourg, I.C.}, \bibinfo{author}{DePaolo, D.J.}, \bibinfo{year}{2012}.
\newblock \bibinfo{title}{Ion desolvation as a mechanism for kinetic isotope fractionation in aqueous systems}.
\newblock \bibinfo{journal}{Proceedings of the National Academy of Sciences} \bibinfo{volume}{109}, \bibinfo{pages}{18689--18694}.
\bibitem[{Ivanov et~al.(2014)Ivanov, Fedorov, Baranchikov and Osiko}]{ivanov2014oriented}
\bibinfo{author}{Ivanov, V.K.}, \bibinfo{author}{Fedorov, P.P.}, \bibinfo{author}{Baranchikov, A.Y.}, \bibinfo{author}{Osiko, V.V.}, \bibinfo{year}{2014}.
\newblock \bibinfo{title}{Oriented attachment of particles: 100 years of investigations of non-classical crystal growth}.
\newblock \bibinfo{journal}{Russian Chemical Reviews} \bibinfo{volume}{83}, \bibinfo{pages}{1204}.
\bibitem[{Jacobson and Holmden(2008)}]{jacobson2008delta44ca}
\bibinfo{author}{Jacobson, A.D.}, \bibinfo{author}{Holmden, C.}, \bibinfo{year}{2008}.
\newblock \bibinfo{title}{{$\delta^{44}$Ca} evolution in a carbonate aquifer and its bearing on the equilibrium isotope fractionation factor for calcite}.
\newblock \bibinfo{journal}{Earth and Planetary Science Letters} \bibinfo{volume}{270}, \bibinfo{pages}{349--353}.
\bibitem[{Jia et~al.(2022)Jia, Zhang, Lammers, Huang and Wang}]{jia2022model}
\bibinfo{author}{Jia, Q.}, \bibinfo{author}{Zhang, S.}, \bibinfo{author}{Lammers, L.}, \bibinfo{author}{Huang, Y.}, \bibinfo{author}{Wang, G.}, \bibinfo{year}{2022}.
\newblock \bibinfo{title}{A model for {pH} dependent strontium partitioning during calcite precipitation from aqueous solutions}.
\newblock \bibinfo{journal}{Chemical Geology} , \bibinfo{pages}{121042}.
\bibitem[{K{\i}sak{\"u}rek et~al.(2011)K{\i}sak{\"u}rek, Eisenhauer, B{\"o}hm, Hathorne and Erez}]{kisakurek2011controls}
\bibinfo{author}{K{\i}sak{\"u}rek, B.}, \bibinfo{author}{Eisenhauer, A.}, \bibinfo{author}{B{\"o}hm, F.}, \bibinfo{author}{Hathorne, E.C.}, \bibinfo{author}{Erez, J.}, \bibinfo{year}{2011}.
\newblock \bibinfo{title}{Controls on calcium isotope fractionation in cultured planktic foraminifera, globigerinoides ruber and globigerinella siphonifera}.
\newblock \bibinfo{journal}{Geochimica et Cosmochimica Acta} \bibinfo{volume}{75}, \bibinfo{pages}{427--443}.
\bibitem[{Krabbenh{\"o}ft et~al.(2010)Krabbenh{\"o}ft, Eisenhauer, B{\"o}hm, Vollstaedt, Fietzke, Liebetrau, Augustin, Peucker-Ehrenbrink, M{\"u}ller, Horn et~al.}]{krabbenhoft2010constraining}
\bibinfo{author}{Krabbenh{\"o}ft, A.}, \bibinfo{author}{Eisenhauer, A.}, \bibinfo{author}{B{\"o}hm, F.}, \bibinfo{author}{Vollstaedt, H.}, \bibinfo{author}{Fietzke, J.}, \bibinfo{author}{Liebetrau, V.}, \bibinfo{author}{Augustin, N.}, \bibinfo{author}{Peucker-Ehrenbrink, B.}, \bibinfo{author}{M{\"u}ller, M.}, \bibinfo{author}{Horn, C.}, et~al., \bibinfo{year}{2010}.
\newblock \bibinfo{title}{Constraining the marine strontium budget with natural strontium isotope fractionations ($\mathrm{^{87}Sr/^{86}Sr^*}$, $\delta\mathrm{^{88/86}Sr}$) of carbonates, hydrothermal solutions and river waters}.
\newblock \bibinfo{journal}{Geochimica et Cosmochimica Acta} \bibinfo{volume}{74}, \bibinfo{pages}{4097--4109}.
\bibitem[{Lammers and Koishi(2021)}]{Lammers2021isotopic}
\bibinfo{author}{Lammers, L.N.}, \bibinfo{author}{Koishi, A.}, \bibinfo{year}{2021}.
\newblock \bibinfo{title}{Isotopic tracers of nonclassical crystallization}, in: \bibinfo{booktitle}{Crystallization via Nonclassical Pathways Volume 2: Aggregation, Biomineralization, Imaging \& Application}. \bibinfo{publisher}{ACS Publications}, pp. \bibinfo{pages}{167--198}.
\bibitem[{Lemarchand et~al.(2004)Lemarchand, Wasserburg and Papanastassiou}]{lemarchand2004rate}
\bibinfo{author}{Lemarchand, D.}, \bibinfo{author}{Wasserburg, G.}, \bibinfo{author}{Papanastassiou, D.}, \bibinfo{year}{2004}.
\newblock \bibinfo{title}{Rate-controlled calcium isotope fractionation in synthetic calcite}.
\newblock \bibinfo{journal}{Geochimica et cosmochimica acta} \bibinfo{volume}{68}, \bibinfo{pages}{4665--4678}.
\bibitem[{Li et~al.(2012)Li, Nielsen, Lee, Frandsen, Banfield and De~Yoreo}]{li2012direction}
\bibinfo{author}{Li, D.}, \bibinfo{author}{Nielsen, M.H.}, \bibinfo{author}{Lee, J.R.}, \bibinfo{author}{Frandsen, C.}, \bibinfo{author}{Banfield, J.F.}, \bibinfo{author}{De~Yoreo, J.J.}, \bibinfo{year}{2012}.
\newblock \bibinfo{title}{Direction-specific interactions control crystal growth by oriented attachment}.
\newblock \bibinfo{journal}{science} \bibinfo{volume}{336}, \bibinfo{pages}{1014--1018}.
\bibitem[{Lorens(1981)}]{lorens1981sr}
\bibinfo{author}{Lorens, R.B.}, \bibinfo{year}{1981}.
\newblock \bibinfo{title}{{Sr, Cd, Mn and Co} distribution coefficients in calcite as a function of calcite precipitation rate}.
\newblock \bibinfo{journal}{Geochimica et Cosmochimica Acta} \bibinfo{volume}{45}, \bibinfo{pages}{553--561}.
\bibitem[{Lupulescu and Rimer(2014)}]{lupulescu2014situ}
\bibinfo{author}{Lupulescu, A.I.}, \bibinfo{author}{Rimer, J.D.}, \bibinfo{year}{2014}.
\newblock \bibinfo{title}{In situ imaging of silicalite-1 surface growth reveals the mechanism of crystallization}.
\newblock \bibinfo{journal}{Science} \bibinfo{volume}{344}, \bibinfo{pages}{729--732}.
\bibitem[{Mavromatis et~al.(2013)Mavromatis, Gautier, Bosc and Schott}]{mavromatis2013kinetics}
\bibinfo{author}{Mavromatis, V.}, \bibinfo{author}{Gautier, Q.}, \bibinfo{author}{Bosc, O.}, \bibinfo{author}{Schott, J.}, \bibinfo{year}{2013}.
\newblock \bibinfo{title}{Kinetics of {Mg} partition and {Mg} stable isotope fractionation during its incorporation in calcite}.
\newblock \bibinfo{journal}{Geochimica et Cosmochimica Acta} \bibinfo{volume}{114}, \bibinfo{pages}{188--203}.
\bibitem[{Mavromatis et~al.(2020)Mavromatis, van Zuilen, Blanchard, van Zuilen, Dietzel and Schott}]{mavromatis2020experimental}
\bibinfo{author}{Mavromatis, V.}, \bibinfo{author}{van Zuilen, K.}, \bibinfo{author}{Blanchard, M.}, \bibinfo{author}{van Zuilen, M.}, \bibinfo{author}{Dietzel, M.}, \bibinfo{author}{Schott, J.}, \bibinfo{year}{2020}.
\newblock \bibinfo{title}{Experimental and theoretical modelling of kinetic and equilibrium {Ba} isotope fractionation during calcite and aragonite precipitation}.
\newblock \bibinfo{journal}{Geochimica et Cosmochimica Acta} \bibinfo{volume}{269}, \bibinfo{pages}{566--580}.
\bibitem[{Michel et~al.(2010)Michel, Barr{\'o}n, Torrent, Morales, Serna, Boily, Liu, Ambrosini, Cismasu and Brown~Jr}]{michel2010ordered}
\bibinfo{author}{Michel, F.M.}, \bibinfo{author}{Barr{\'o}n, V.}, \bibinfo{author}{Torrent, J.}, \bibinfo{author}{Morales, M.P.}, \bibinfo{author}{Serna, C.J.}, \bibinfo{author}{Boily, J.F.}, \bibinfo{author}{Liu, Q.}, \bibinfo{author}{Ambrosini, A.}, \bibinfo{author}{Cismasu, A.C.}, \bibinfo{author}{Brown~Jr, G.E.}, \bibinfo{year}{2010}.
\newblock \bibinfo{title}{Ordered ferrimagnetic form of ferrihydrite reveals links among structure, composition, and magnetism}.
\newblock \bibinfo{journal}{Proceedings of the National Academy of Sciences} \bibinfo{volume}{107}, \bibinfo{pages}{2787--2792}.
\bibitem[{Mills et~al.(2021)Mills, DePaolo and Lammers}]{mills2021influence}
\bibinfo{author}{Mills, J.V.}, \bibinfo{author}{DePaolo, D.J.}, \bibinfo{author}{Lammers, L.N.}, \bibinfo{year}{2021}.
\newblock \bibinfo{title}{The influence of {Ca}:{CO$_3$} stoichiometry on ca isotope fractionation: Implications for process-based models of calcite growth}.
\newblock \bibinfo{journal}{Geochimica et Cosmochimica Acta} \bibinfo{volume}{298}, \bibinfo{pages}{87--111}.
\bibitem[{Nehrke et~al.(2007)Nehrke, Reichart, Van~Cappellen, Meile and Bijma}]{nehrke2007dependence}
\bibinfo{author}{Nehrke, G.}, \bibinfo{author}{Reichart, G.J.}, \bibinfo{author}{Van~Cappellen, P.}, \bibinfo{author}{Meile, C.}, \bibinfo{author}{Bijma, J.}, \bibinfo{year}{2007}.
\newblock \bibinfo{title}{Dependence of calcite growth rate and {Sr} partitioning on solution stoichiometry: Non-kossel crystal growth}.
\newblock \bibinfo{journal}{Geochimica et Cosmochimica Acta} \bibinfo{volume}{71}, \bibinfo{pages}{2240--2249}.
\bibitem[{Nielsen et~al.(2013)Nielsen, De~Yoreo and DePaolo}]{nielsen2013general}
\bibinfo{author}{Nielsen, L.C.}, \bibinfo{author}{De~Yoreo, J.J.}, \bibinfo{author}{DePaolo, D.J.}, \bibinfo{year}{2013}.
\newblock \bibinfo{title}{General model for calcite growth kinetics in the presence of impurity ions}.
\newblock \bibinfo{journal}{Geochimica et Cosmochimica Acta} \bibinfo{volume}{115}, \bibinfo{pages}{100--114}.
\bibitem[{Nielsen et~al.(2012)Nielsen, DePaolo and De~Yoreo}]{nielsen2012self}
\bibinfo{author}{Nielsen, L.C.}, \bibinfo{author}{DePaolo, D.J.}, \bibinfo{author}{De~Yoreo, J.J.}, \bibinfo{year}{2012}.
\newblock \bibinfo{title}{Self-consistent ion-by-ion growth model for kinetic isotopic fractionation during calcite precipitation}.
\newblock \bibinfo{journal}{Geochimica et Cosmochimica Acta} \bibinfo{volume}{86}, \bibinfo{pages}{166--181}.
\bibitem[{Nielsen et~al.(2014)Nielsen, Aloni and De~Yoreo}]{nielsen2014situ}
\bibinfo{author}{Nielsen, M.H.}, \bibinfo{author}{Aloni, S.}, \bibinfo{author}{De~Yoreo, J.J.}, \bibinfo{year}{2014}.
\newblock \bibinfo{title}{In situ {TEM} imaging of {CaCO$_3$} nucleation reveals coexistence of direct and indirect pathways}.
\newblock \bibinfo{journal}{Science} \bibinfo{volume}{345}, \bibinfo{pages}{1158--1162}.
\bibitem[{Paquette and Reeder(1995)}]{paquette1995relationship}
\bibinfo{author}{Paquette, J.}, \bibinfo{author}{Reeder, R.J.}, \bibinfo{year}{1995}.
\newblock \bibinfo{title}{Relationship between surface structure, growth mechanism, and trace element incorporation in calcite}.
\newblock \bibinfo{journal}{Geochimica et Cosmochimica Acta} \bibinfo{volume}{59}, \bibinfo{pages}{735--749}.
\bibitem[{Pokrovsky and Schott(2002)}]{pokrovsky2002surface}
\bibinfo{author}{Pokrovsky, O.}, \bibinfo{author}{Schott, J.}, \bibinfo{year}{2002}.
\newblock \bibinfo{title}{Surface chemistry and dissolution kinetics of divalent metal carbonates}.
\newblock \bibinfo{journal}{Environmental science \& technology} \bibinfo{volume}{36}, \bibinfo{pages}{426--432}.
\bibitem[{Putnis et~al.(2021)Putnis, Wang, Ruiz-Agudo, Ruiz-Agudo and Renard}]{putnis2021crystallization}
\bibinfo{author}{Putnis, C.V.}, \bibinfo{author}{Wang, L.}, \bibinfo{author}{Ruiz-Agudo, E.}, \bibinfo{author}{Ruiz-Agudo, C.}, \bibinfo{author}{Renard, F.}, \bibinfo{year}{2021}.
\newblock \bibinfo{title}{Crystallization via nonclassical pathways: Nanoscale imaging of mineral surfaces}, in: \bibinfo{booktitle}{Crystallization via Nonclassical Pathways Volume 2: Aggregation, Biomineralization, Imaging \& Application}. \bibinfo{publisher}{ACS Publications}, pp. \bibinfo{pages}{1--35}.
\bibitem[{Shao et~al.(2021)Shao, Farka{\v{s}}, Mosley, Tyler, Wong, Chamberlayne, Raven, Samanta, Holmden, Gillanders et~al.}]{shao2021impact}
\bibinfo{author}{Shao, Y.}, \bibinfo{author}{Farka{\v{s}}, J.}, \bibinfo{author}{Mosley, L.}, \bibinfo{author}{Tyler, J.}, \bibinfo{author}{Wong, H.}, \bibinfo{author}{Chamberlayne, B.}, \bibinfo{author}{Raven, M.}, \bibinfo{author}{Samanta, M.}, \bibinfo{author}{Holmden, C.}, \bibinfo{author}{Gillanders, B.M.}, et~al., \bibinfo{year}{2021}.
\newblock \bibinfo{title}{Impact of salinity and carbonate saturation on stable {Sr} isotopes ({$\delta^{88/86}$Sr}) in a lagoon-estuarine system}.
\newblock \bibinfo{journal}{Geochimica et Cosmochimica Acta} \bibinfo{volume}{293}, \bibinfo{pages}{461--476}.
\bibitem[{Stevenson et~al.(2014)Stevenson, Hermoso, Rickaby, Tyler, Minoletti, Parkinson, Mokadem and Burton}]{stevenson2014controls}
\bibinfo{author}{Stevenson, E.I.}, \bibinfo{author}{Hermoso, M.}, \bibinfo{author}{Rickaby, R.E.}, \bibinfo{author}{Tyler, J.J.}, \bibinfo{author}{Minoletti, F.}, \bibinfo{author}{Parkinson, I.J.}, \bibinfo{author}{Mokadem, F.}, \bibinfo{author}{Burton, K.W.}, \bibinfo{year}{2014}.
\newblock \bibinfo{title}{Controls on stable strontium isotope fractionation in coccolithophores with implications for the marine sr cycle}.
\newblock \bibinfo{journal}{Geochimica et cosmochimica acta} \bibinfo{volume}{128}, \bibinfo{pages}{225--235}.
\bibitem[{Tang et~al.(2008a)Tang, Dietzel, B{\"o}hm, K{\"o}hler and Eisenhauer}]{tang2008II}
\bibinfo{author}{Tang, J.}, \bibinfo{author}{Dietzel, M.}, \bibinfo{author}{B{\"o}hm, F.}, \bibinfo{author}{K{\"o}hler, S.J.}, \bibinfo{author}{Eisenhauer, A.}, \bibinfo{year}{2008}a.
\newblock \bibinfo{title}{{Sr}$^{2+}$/{Ca}$^{2+}$ and $^{44}${Ca}/$^{40}${Ca} fractionation during inorganic calcite formation: {II}. {Ca} isotopes}.
\newblock \bibinfo{journal}{Geochimica et Cosmochimica Acta} \bibinfo{volume}{72}, \bibinfo{pages}{3733--3745}.
\bibitem[{Tang et~al.(2008b)Tang, K{\"o}hler and Dietzel}]{tang2008I}
\bibinfo{author}{Tang, J.}, \bibinfo{author}{K{\"o}hler, S.J.}, \bibinfo{author}{Dietzel, M.}, \bibinfo{year}{2008}b.
\newblock \bibinfo{title}{{Sr}$^{2+}$/{Ca}$^{2+}$ and $^{44}${Ca}/$^{40}${Ca} fractionation during inorganic calcite formation: {I}. {Sr} incorporation}.
\newblock \bibinfo{journal}{Geochimica et Cosmochimica Acta} \bibinfo{volume}{72}, \bibinfo{pages}{3718--3732}.
\bibitem[{Teng et~al.(2000)Teng, Dove and De~Yoreo}]{teng2000kinetics}
\bibinfo{author}{Teng, H.H.}, \bibinfo{author}{Dove, P.M.}, \bibinfo{author}{De~Yoreo, J.J.}, \bibinfo{year}{2000}.
\newblock \bibinfo{title}{Kinetics of calcite growth: surface processes and relationships to macroscopic rate laws}.
\newblock \bibinfo{journal}{Geochimica et Cosmochimica Acta} \bibinfo{volume}{64}, \bibinfo{pages}{2255--2266}.
\bibitem[{Tesoriero and Pankow(1996)}]{tesoriero1996solid}
\bibinfo{author}{Tesoriero, A.J.}, \bibinfo{author}{Pankow, J.F.}, \bibinfo{year}{1996}.
\newblock \bibinfo{title}{Solid solution partitioning of {Sr$^{2+}$}, {Ba$^{2+}$}, and {Cd$^{2+}$} to calcite}.
\newblock \bibinfo{journal}{Geochimica et Cosmochimica Acta} \bibinfo{volume}{60}, \bibinfo{pages}{1053--1063}.
\bibitem[{Voigt et~al.(2015)Voigt, Hathorne, Frank, Vollstaedt and Eisenhauer}]{voigt2015variability}
\bibinfo{author}{Voigt, J.}, \bibinfo{author}{Hathorne, E.C.}, \bibinfo{author}{Frank, M.}, \bibinfo{author}{Vollstaedt, H.}, \bibinfo{author}{Eisenhauer, A.}, \bibinfo{year}{2015}.
\newblock \bibinfo{title}{Variability of carbonate diagenesis in equatorial pacific sediments deduced from radiogenic and stable {Sr} isotopes}.
\newblock \bibinfo{journal}{Geochimica et cosmochimica acta} \bibinfo{volume}{148}, \bibinfo{pages}{360--377}.
\bibitem[{Wang et~al.(2023a)Wang, Di, Asael, Planavsky and Tarhan}]{wang2023investigation}
\bibinfo{author}{Wang, J.}, \bibinfo{author}{Di, Y.}, \bibinfo{author}{Asael, D.}, \bibinfo{author}{Planavsky, N.J.}, \bibinfo{author}{Tarhan, L.G.}, \bibinfo{year}{2023}a.
\newblock \bibinfo{title}{An investigation of factors affecting high-precision {Sr} isotope analyses ({$^{87}$Sr/$^{86}$Sr} and {$\delta^{88/86}$Sr)} by {MC-ICP-MS}}.
\newblock \bibinfo{journal}{Chemical Geology} \bibinfo{volume}{621}, \bibinfo{pages}{121365}.
\bibitem[{Wang et~al.(2021)Wang, Jacobson, Sageman and Hurtgen}]{wang2021stable}
\bibinfo{author}{Wang, J.}, \bibinfo{author}{Jacobson, A.D.}, \bibinfo{author}{Sageman, B.B.}, \bibinfo{author}{Hurtgen, M.T.}, \bibinfo{year}{2021}.
\newblock \bibinfo{title}{Stable {Ca} and {Sr} isotopes support volcanically triggered biocalcification crisis during oceanic anoxic event 1a}.
\newblock \bibinfo{journal}{Geology} \bibinfo{volume}{49}, \bibinfo{pages}{515--519}.
\bibitem[{Wang et~al.(2023b)Wang, Jacobson, Sageman and Hurtgen}]{wang2023application}
\bibinfo{author}{Wang, J.}, \bibinfo{author}{Jacobson, A.D.}, \bibinfo{author}{Sageman, B.B.}, \bibinfo{author}{Hurtgen, M.T.}, \bibinfo{year}{2023}b.
\newblock \bibinfo{title}{Application of the {$\delta^{44/40}$Ca}--{$\delta^{88/86}$Sr} multi-proxy to namibian marinoan cap carbonates}.
\newblock \bibinfo{journal}{Geochimica et Cosmochimica Acta} .
\bibitem[{Watkins et~al.(2017)Watkins, DePaolo and Watson}]{watkins2017kinetic}
\bibinfo{author}{Watkins, J.M.}, \bibinfo{author}{DePaolo, D.J.}, \bibinfo{author}{Watson, E.B.}, \bibinfo{year}{2017}.
\newblock \bibinfo{title}{Kinetic fractionation of non-traditional stable isotopes by diffusion and crystal growth reactions}.
\newblock \bibinfo{journal}{Reviews in Mineralogy and Geochemistry} \bibinfo{volume}{82}, \bibinfo{pages}{85--125}.
\bibitem[{Watson(2004)}]{watson2004conceptual}
\bibinfo{author}{Watson, E.B.}, \bibinfo{year}{2004}.
\newblock \bibinfo{title}{A conceptual model for near-surface kinetic controls on the trace-element and stable isotope composition of abiogenic calcite crystals}.
\newblock \bibinfo{journal}{Geochimica et cosmochimica acta} \bibinfo{volume}{68}, \bibinfo{pages}{1473--1488}.
\bibitem[{Zachara et~al.(1991)Zachara, Cowan and Resch}]{zachara1991sorption}
\bibinfo{author}{Zachara, J.}, \bibinfo{author}{Cowan, C.}, \bibinfo{author}{Resch, C.}, \bibinfo{year}{1991}.
\newblock \bibinfo{title}{Sorption of divalent metals on calcite}.
\newblock \bibinfo{journal}{Geochimica et cosmochimica acta} \bibinfo{volume}{55}, \bibinfo{pages}{1549--1562}.
\bibitem[{Zhang and DePaolo(2020)}]{zhang2020equilibrium}
\bibinfo{author}{Zhang, S.}, \bibinfo{author}{DePaolo, D.J.}, \bibinfo{year}{2020}.
\newblock \bibinfo{title}{Equilibrium calcite-fluid {Sr/Ca} partition coefficient from marine sediment and pore fluids}.
\newblock \bibinfo{journal}{Geochimica et Cosmochimica Acta} \bibinfo{volume}{289}, \bibinfo{pages}{33--46}.

\end{thebibliography}

\end{document}